\documentclass[reprint,amsmath,amssymb,aps,prb,]{revtex4-2}
\usepackage{graphicx}
\usepackage{dcolumn}
\usepackage{bm}
\usepackage{hyperref}

\usepackage{amssymb,amsthm}
\usepackage{braket}
\usepackage{dsfont}
\usepackage{xcolor}

\begin{document}
\title{Stability of the Quantum Coherent Superradiant States in Relation to
Exciton-Phonon Interactions and the Fundamental Soliton in Hybrid Perovskites}

\author{A. A. Gladkij}
\affiliation{Ioffe Physical-Technical Institute, Russian Academy of Sciences, St.
Petersburg 194021, Russia}
\author{N. A. Veretenov}
\affiliation{Ioffe Physical-Technical Institute, Russian Academy of Sciences, St.
Petersburg 194021, Russia}
\author{N. N. Rosanov}
\affiliation{Ioffe Physical-Technical Institute, Russian Academy of Sciences, St.
Petersburg 194021, Russia}
\author{B. A. Malomed}
\affiliation{Department of Physical Electronics, School of Electrical Engineering, Faculty
of Engineering, and Center for Light-Matter Interaction, Tel Aviv University,
P.O.B. 39040, Tel Aviv, Israel}
\affiliation{Instituto de Alta Investigacion, Universidad de Tarapaca, Casilla 7D, Arica, Chile}
\author{V. Al. Osipov}
\affiliation{Institute for Advanced Study in Mathematics, Harbin Institute of Technology,
92 West Da Zhi Street, Harbin, 150001, China}
\affiliation{Suzhou Research Institute, Harbin Institute of Technology, 500 South Guandu
Road, Suzhou, 215104, China}
\author{B. D. Fainberg}
\email{fainberg@hit.ac.il}
\affiliation{Faculty of Sciences, Physics Department, Holon Institute of Technology, 52
Golomb Street, POB 305, Holon 5810201, Israel}

\date{\today }

\begin{abstract}
The use of macroscopic coherent quantum states at room temperature is crucial
in modern quantum technologies. In light of recent experiments demonstrating
high-temperature superfluorescence in hybrid perovskite thin films, in this
work we investigate the stability of the superradiant state concerning
exciton-phonon interactions, taking into account the specifics of perovskites.
We focused on quasi-2D Wannier excitons
interacting with longitudinal optical (LO) phonons in polar crystals, as well
as with acoustic phonons. Our study leads to the derivation of nonlinear
equations in the coordinate space that govern the exciton wavefunction's
coefficient in the single-exciton basis for the lowest exciton state, which
translates to the complex-valued polarization. The resulting equations take
the form of a 2D nonlocal nonlinear Schrodinger (NLS) equation. We perform a
linear stability analysis of the plane wave solutions for the equations in
question, which allows us to establish stability criteria. This analysis is
particularly important for evaluating the stability of the superradiant state
in the considered quasi-2D structures, as the superradiant state represents a
specific case of the plane wave solution. Our findings indicate that, when the
exciton interacts with LO phonons, a plane wave solution is modulationally
stable, provided that the square of its amplitude does not exceed a critical
intensity value defined by the exciton-LO phonon interaction parameters. In
scenarios involving the weakly nonlocal NLS equation, we find that it
transitions into a purely nonlocal form. The linear stability analysis yields
results that align with those derived from the equation that does not consider
weak non-locality in the long-wave limit. Furthermore, interactions between
the exciton and acoustic phonons reduce the intensity of modulationally stable
waves compared to the case without such interactions. Our analytical results
are corroborated by numerical calculations. We also numerically solve the 2D
nonlocal NLS equation in the polar coordinates and obtain its fundamental
soliton solution, which is stable.
\end{abstract}

\maketitle

\section{Introduction}
The use of room-temperature macroscopic quantum coherence is one of the most
exciting frontiers in modern physics and materials science, because it opens
the door to applying quantum effects in realistic, everyday conditions without
the need for cryogenics. Recently, superfluorescence (SF) based on the
macroscopic quantum coherence was observed in methyl ammonium lead iodide
($MAPbI_{3}$ thin film) at a temperature of 78 K
\cite{Gundogdu2021Nature_Phot}, and then in quasi-2D structures such as
phenethylammonium cesium lead bromide ($PEA:CsPbBr_{3}$)
\cite{Gundogdu2022Nature_Phot} and the superlattice (crystal) structure of the
perovskite ($PEA_{2}(FAPbBr_{3})_{n-1}PbBr_{4}$)
\cite{Wildenborg2025ACSPhotonics} at a room temperature. The discovery of
high-temperature SF in hybrid perovskites naturally raised questions regarding
the underlying mechanism that was elucidated in
Ref.\cite{Fainberg_Osipov2024JCP}. The physical picture of the
high-temperature SF drawn from the analysis of the experiments conducted by
Biliroglu \textit{et al.} al.~\cite{Gundogdu2022Nature_Phot} suggests that the
formation of large polarons protects the electronic excitation from dephasing
even at room temperature. The model of Ref.\cite{Fainberg_Osipov2024JCP}
describes a quasi-2D Wannier exciton in a thin film interacting with phonons
via the longitudinal optical (LO) phonon-exciton Frohlich interaction. Using
the multiconfiguration Hartree approach, equations of motion for a
single-exciton wavefunction in the momentum (wave number $\mathbf{q}$) space
were derived, where the vibrational degrees of freedom interact with the
Wannier exciton through a mean-field Hartree term. Following the physical
derivation of the equations step by step, we can see that the class of
admissible polar crystals becomes progressively restricted. This procedure
ultimately leads to Wannier excitons in quasi-2D perovskite thin films,
which uniquely possess the required properties. Specifically, Ref.\cite%
{Fainberg_Osipov2024JCP} hast demonstrated that the radiating system must be a
Wannier exciton, and therefore its binding energy must substantially exceed
the thermal energy at room temperature that is realized for quasi-2D
perovskites \cite{Long19Nanoscale}. In addition, the equations derived in
the present paper rely on the existence of low-energy, long-lived LO phonon
modes that couple to electronic transitions in $CsPbBr_{3}$ \cite{Miyata17},
with lifetimes on the order of tens of picoseconds. Such long-lived LO
phonon modes may be associated with lead-halide-lead rocking vibrations
\cite{Gundogdu2025Nature}.

The analysis of small
deviations from the coherent superradiant state ($\mathbf{q}=0$)conducted
in Ref.\cite{Fainberg_Osipov2024JCP} showed that
the "vibrational" contribution to the superradiant state attenuation was
absent up to the order of quadratic $\sim k^{2}$ terms with respect to the
wave number. Such a behavior arises because for $k=0$, the macroscopic
electric field of the LO phonon is uniform in space. The exciton is neutral,
so its energy cannot be changed by a uniform field. For small non-zero $k$,
the electronic matrix element of the Wannier exciton-LO phonon interaction,
$D_{l}^{pol}(\mathbf{k};00)$, is proportional to $k$, analogous to the matrix
element for optical quadrupole transitions. Therefore a state with a specific
wave vector $\mathbf{q}$ (including $\mathbf{q}=0$) behaves like metastable
state in optics when there is no dipole transition. The consideration of
Ref.\cite{Fainberg_Osipov2024JCP} drew an analogy between the system under
study and stable magnetic systems, where the conservation laws determined the
nullification of the constant (momentum-independent) decay rate part. In the
exciton-phonon system, the nullification is associated with the absence of a
momentum-independent component in the Wannier exciton-LO phonon interaction
coupling function $D_{l}^{pol}(\mathbf{k};00)$.

Nonetheless, several unresolved issues remain. First, there is the question of
how LO phonon-exciton Frohlich interactions influence the superradiant
characteristics at length scales comparable to or smaller than $2\pi a_{0}$,
where $a_{0}$ is the exciton radius. This extends beyond the quadratic $\sim
k^{2}$ expansion near the superradiant state used in
Ref.\cite{Fainberg_Osipov2024JCP}. Second, the stability of the superradiant
state presents another challenge. The core difficulty lies in the fact that
the equations derived within the Hartree framework are nonlinear
\cite{Fainberg_Osipov2024JCP} and can be reduced to the 2D nonlocal nonlinear
Schrodinger (NLS) equation \cite{Malomed2022} in the coordinate space (as
further discussed below). It is well-known that solutions to such equations
may exhibit modulational instability (MI)
\cite{Krolikowski2001,Krolikowsk2004}. One of the goals of this study is to
identify the critical squared amplitude of a single-exciton wavefunction at
which the superradiant state remains stable. This critical value is expected
to depend on the strength of the exciton-phonon interaction. Third, there are
indications that it might be worth going beyond the Frohlich framework and
account for the interplay between short-range (exciton-acoustic phonon) and
long-range (exciton-LO phonon) coupling in the systems under investigation
\cite{Silva2020JPCL}.

To address the problems mentioned above, we transition from equations in the
momentum space to those in the coordinate space, as described in Section
\ref{Section:real space}. In Section \ref{Section:solution1}, we present
solutions to these equations in the form of plane waves, with the superradiant
state appearing as a special case, and perform their linear stability analysis
including numerical results. Section \ref{Section:weakly nonlocal} examines
the limit of the weakly nonlocal NLS equation. In
Section\ref{Section:acoustic} we move beyond the Frohlich picture and
investigate the interplay between exciton-acoustic phonon and exciton-LO
phonon lattice couplings. In Section\ref{Section:fundamental soliton} we
numerically solve the 2D nonlocal NLS equation in the polar coordinates and
obtain its fundamental soliton solution. Finally, Section
\ref{Section:conclusion}, provides a brief conclusion.

\section{The basic equation in the coordinate space}

\label{Section:real space}

We consider a quasi-2D Wannier exciton interacting with LO phonons in a polar
crystal \cite{Fainberg_Osipov2024JCP}, bearing in mind two energy bands: the
valence band ($v$) and the conductance band ($c$). To introduce the
single-exciton wavefunction, we introduce the basis set of single-exciton
states $|n\mathbf{q}\rangle=B_{n\mathbf{q}}^{\dag}|0\rangle$ with energies
$\hbar W_{n}(\mathbf{k})$ where $B_{n\mathbf{q}}^{\dag}$ are creation
operators of excitons, $\mathbf{q}$ is a wave-vector, and $n$ is the principal
quantum number of an excitonic state ($n=0,1,2$ ... for 2D). $\hbar
W_{n}(\mathbf{k})=E_{c}(\mathbf{k}_{0})-E_{v}(\mathbf{k}_{0})+E_{n}%
+\frac{\hbar^{2}k^{2}}{2M^{\ast}}$ where $E_{n}$ is the energy of the internal
structure of an exciton in state $n$, $E_{c}(\mathbf{k}_{0})-E_{v}%
(\mathbf{k}_{0})$ is the energy gap between the minimum of the first
conduction band and the maximum of the valence band, $M^{\ast}=m_{e}^{\ast
}+m_{h}^{\ast}$, $m_{e}^{\ast}$ and $m_{h}^{\ast}$ is the electron and hole
effective mass, respectively. The exciton wavefunction is therefore given by
$|\Psi\rangle=\sum_{\mathbf{q}}[G(\mathbf{q}|t)|0\rangle+\sum_{n}%
C_{n}(\mathbf{q}|t)|n\mathbf{q}\rangle]$, where $G(\mathbf{q}|t)$ is the
ground state amplitude, and the coefficients $C_{n}(\mathbf{q}|t)$ represent
the exciton wavefunction in the single-exciton basis. The complete
single-exciton wave function satisfies the normalization condition
\begin{equation}
\langle\Psi(t)|\Psi(t)\rangle=\sum_{\mathbf{q}}[|G(\mathbf{q}|t)|^{2}+\sum
_{n}|C_{n}(\mathbf{q}|t)|^{2}]=1. \label{eq:Normalization}%
\end{equation}

We use the coherent states $|\sigma\rangle$ \cite{Glauber63,BBGK1971} as the
basis for the phonon states. Each coherent state $|\sigma\rangle$ is
parametrized by a multidimensional complex-valued vector $\sigma$ that encodes
the coherent state center, i.e. the classical coordinate $x$ and classical
momentum $p$, namely $\sigma=x+ip$. Thus, our working basis consists of the direct
products of the exciton and vibrational states%
\begin{equation}
|\mathbf{\sigma},n\mathbf{q}\rangle=|\mathbf{\sigma}\rangle|n\mathbf{q}%
\rangle,\quad|\mathbf{\sigma}\rangle=\bigotimes_{\mathbf{q}}|\sigma
_{\mathbf{q}}\rangle. \label{ketsigmaa}%
\end{equation}

To describe the time evolution of the wavefunction $|\Psi(t)\rangle$, we use
the Ansatz for the time-dependent basis vectors (see also Davydov
Ansatz~\cite{Davydov82,Gelin2015} used in the theory of one-dimensional
molecular aggregates). It assumes that the basis vectors $|\mathbf{\sigma
}(t),n\mathbf{q}\rangle$ depend on time in addition to the time-dependent
expansion coefficients $C_{n}(\mathbf{\sigma},\mathbf{q}|t)\equiv
C_{n}(\mathbf{q}|t)$. With this, the wavefunction can be expanded as
\begin{equation}
|\Psi(t)\rangle=\sum_{\mathbf{\sigma}}\sum_{n\mathbf{q}}C_{n}(\mathbf{\sigma
},\mathbf{q}|t)|\mathbf{\sigma}(t),n\mathbf{q}\rangle+\sum_{\mathbf{q}%
}G(\mathbf{q}|t)|0\rangle. \label{wavefunction1}%
\end{equation}

It is convenient to formulate the equations in terms of the complex-valued
polarization
\begin{equation}
Q_{n}(\mathbf{q}|t)\equiv\langle\Psi(t)|\hat{B}_{n\mathbf{q}}|\Psi
(t)\rangle=\sum_{\mathbf{q}^{\prime}}G^{\ast}(\mathbf{q}^{\prime}%
|t)C_{n}(\mathbf{q}|t). \label{eq:Q_n(q|t)}%
\end{equation}
In the first approximation, when the population of the ground state is close
to $1$, one can set $\sum_{\mathbf{q}^{\prime}}G^{\ast}(\mathbf{q}^{\prime
}|t)\approx1$, and $Q_{n}(\mathbf{q}|t)\simeq C_{n}(\mathbf{q}|t)$. In other
words, the complex-valued polarization is close to the coefficient
$C_{n}(\mathbf{q}|t)$.

The electronic matrix element for the coupling function of the Wannier
exciton-LO phonon interaction, $D_{l}^{pol}(\mathbf{k};nn^{\prime })$,
is largest for the lowest exciton state ($n=n^{\prime }=0$), and is given by Eq.(11) of
Ref.\cite{Fainberg_Osipov2024JCP}. Therefore, below we will consider only this
state, i.e. we assume that exciton states other than $n=0$ are not excited. Here the wave vector
$\mathbf{k}$ is related to the
Fourier  transform of the charge distributions of the electron and hole in
the internal exciton motion.
The quantity $D_{l}^{pol}(\mathbf{k};nn^{\prime })$ defines the effectiveness of the exciton interaction with a
particular phonon $\mathbf{k}$.

In this section, our aim is to derive a nonlinear equation governing the
coefficient $C_{0}$, or equivalently, the complex-valued polarization $Q_{0}$,
in the coordinate space. To initiate this process, we combine the first and
second equations of motion formulated in momentum space - namely Eqs. (29) and
(32) of Ref. \cite{Fainberg_Osipov2024JCP} - into a single unified expression.
Equation (32) of Ref.\cite{Fainberg_Osipov2024JCP} for the coefficient
$C_{n}(\mathbf{q}|t)$ when $n=0$ is
\begin{align}
\dot{C}_{0}(\mathbf{\sigma },\mathbf{q}|t)& =-iW_{0}(\mathbf{q})C_{0}(%
\mathbf{\sigma },\mathbf{q}|t)  \notag \\
& +\sum_{\mathbf{k}}\tilde{\alpha}_{l}(\mathbf{k};00)C_{0}(\mathbf{\sigma },%
\mathbf{q}+\mathbf{k}|t)-\hat{\Gamma}(C)  \label{eq:dC/dt}
\end{align}%
where the term $\hat{\Gamma}(C)$ describes spontaneous decay of the state
under consideration. Below, we omit this term, considering it
to be unimportant for the processes considered in the present paper.
Equation (\ref{eq:dC/dt}) differs from Eq.(32) of Ref.\cite%
{Fainberg_Osipov2024JCP} by omitting an insignificant correction to the
exciton eigenenergy $\hbar W_{0}(\mathbf{q})$.The parameter of
electron-vibrational coupling $\tilde{\alpha}_{l}\mathbf{(k};00)=
D_{l}^{pol\ast}\mathbf{(k};00)(\sigma_{\mathbf{k}}^{\ast}-\sigma_{\mathbf{-k}%
})$. The latter is defined by the first equation of motion for the vibration
degrees of freedom - namely for quantity $\sigma_{\mathbf{k}}$, see Eq.(29) of
Ref.\cite{Fainberg_Osipov2024JCP}:%

\begin{equation}
\dot{\sigma}_{\mathbf{k}}=-(i\omega_{l}+\gamma)\sigma_{\mathbf{k}}+D_{l}%
^{pol}(-\mathbf{k};00)F_{00}(\mathbf{k}|t). \label{eq:dsigma_ks/dt}%
\end{equation}
Here the mean-field Hartree term is $F_{00}(\mathbf{k}|t)=\sum_{\mathbf{q}%
}C_{0}^{\ast}(\mathbf{q}|t)C_{0}(\mathbf{q}+\mathbf{k}|t)/\sum_{\mathbf{q}%
^{\prime}}|C_{0}(\mathbf{q}^{\prime}|t)|^{2}$, and $\omega_{l}$ is the low-energy,
long-lived LO phonon mode that couple to electronic
transitions in $CsPbBr_{3}$ \cite{Miyata17}, with lifetimes on the order of
tens of picoseconds. Such a long-lived LO phonon mode may be associated with
lead-halide-lead rocking vibrations \cite{Gundogdu2025Nature}.
If the single excited state is $n=0$, then $\sum
_{\mathbf{q}^{\prime}}|C_{0}(\mathbf{q}^{\prime}|t)|^{2}=1-\sum_{\mathbf{q}%
}|G(\mathbf{q}|t)|^{2}$. An alternative derivation of the equations for
$\sigma_{\mathbf{k}}$ and $C_{n}(\mathbf{q}|t)$ is given in Appendix A.

Using the solution of Eq.(\ref{eq:dsigma_ks/dt}) independent of the initial
conditions, the expression for $\tilde{\alpha}_{l}(\mathbf{k};00)$ can be
written in the following form \cite{Fainberg_Osipov2024JCP}

\begin{eqnarray}
\tilde{\alpha}_{l}(\mathbf{k};00) &=&2i|D_{l}^{pol}(\mathbf{k}%
;00)|^{2}\int_{0}^{t}F_{00}(\mathbf{k}|t-\tau )\exp (-\gamma \tau )
\nonumber \\
&&\times \sin \left( \omega _{l}\tau \right) d\tau .
\label{eq.:alpha^tilda_l(k)2}
\end{eqnarray}

Since the LO frequency $\omega_{l}\ $is about $100$ $cm^{\mathbf{-}1}$ in
physical units, the integral on the right-hand side of
Eq.(\ref{eq.:alpha^tilda_l(k)2}) is a convolution of the Hartree term
$F_{00}(\mathbf{k}|t-\tau)$ with the fast varying function $\exp(-\gamma
\tau)\sin\left(  \omega_{l}\tau\right)  $. Considering $F_{00}(\mathbf{k}%
|t-\tau)$ to be a slower function of time, we get
\begin{equation}
\tilde{\alpha}_{l}(\mathbf{k};00)\approx i2\omega_{0}(k)F_{00}(\mathbf{k}|t)
\label{eq.:alpha^tilda_l(k)2corr_steady}%
\end{equation}
where
\begin{equation}
\omega_{0}(k)=\frac{\omega_{l}}{\omega_{l}^{2}+\gamma^{2}}|D_{l}%
^{pol}(\mathbf{k};00)|^{2}. \label{eq:omega_0(k)}%
\end{equation}

To move to equations in the coorinate space, we perform the Fourier transform
of Eq.(32) for the \ coefficient $C_{n}(\mathbf{q}|t)$ of
Ref.\cite{Fainberg_Osipov2024JCP} in the momentum space, using
\cite{Kit63,Dav71,Agranovich09,Fainberg19JPCC}%
\begin{eqnarray}
C_{0}(\mathbf{q}|t) &=&\frac{1}{\sqrt{N}}\sum_{m}C_{0m}\exp (-i\mathbf{q}%
\mathbf{R}_{m}),  \nonumber \\
C_{0m} &=&\frac{1}{\sqrt{N}}\sum_{\mathbf{q}}C_{0}(\mathbf{q}|t)\exp (i%
\mathbf{q}\mathbf{R}_{m}),  \label{eq:FT}
\end{eqnarray}%
where $N$ is the number of particles (electron-hole pairs)
\cite{Fainberg_Osipov2024JCP}, and $\mathbf{R}_{m}$ is the exciton
center-of-mass coordinate. The Fourier transform of the parameter of the
electron-vibrational coupling $\tilde{\alpha}_{l}\mathbf{(k};00)$ can be
written as

\begin{equation}
\tilde{\alpha}_{l}(\mathbf{k};00)=\frac{1}{N}\sum_{m}\tilde{\alpha}%
_{m}e^{i\mathbf{kR}_{m}}, \quad\tilde{\alpha}_{m}=\sum_{\mathbf{k}}%
\tilde{\alpha}_{l}(\mathbf{k};00)\exp(-i\mathbf{kR}_{m}%
).\label{eq.:alpha^tilda_m}%
\end{equation}
Then Eq.(32) of Ref.\cite{Fainberg_Osipov2024JCP} for $n=0$ takes the form
\begin{eqnarray}
\frac{d}{dt}C_{0m}(t) &=&[-i\bar{W}_{0}-\frac{1}{2}\Gamma _{0}(\omega )+%
\frac{i\hbar }{2M^{\ast }}\frac{\partial ^{2}}{\partial \mathbf{R}_{m}^{2}}
\nonumber \\
&&+2i\sum_{m^{\prime }}|C_{0m^{\prime }}|^{2}\omega _{0}(R_{m^{\prime
}m})]C_{0m}(t),  \label{eq.:dC_om(t)/dt3}
\end{eqnarray}%
which, except for the term $-\frac{1}{2}\Gamma_{0}(\omega)C_{0m}(t)$ on the
right-hand side, is a 2D nonlocal nonlinear Schrodinger (NLS) equation widely
used in the theory of solitons \cite{Krolikowski2000,Malomed2022}. In this
regard, it is worth noting a recent experimental study \cite{Gundogdu2025Nature} in which the soliton concept was used to explain high-temperature SF in perovskites. This provides experimental support for the theoretical considerations presented in our work.

\begin{figure}
\begin{center}
\includegraphics[scale=0.5]{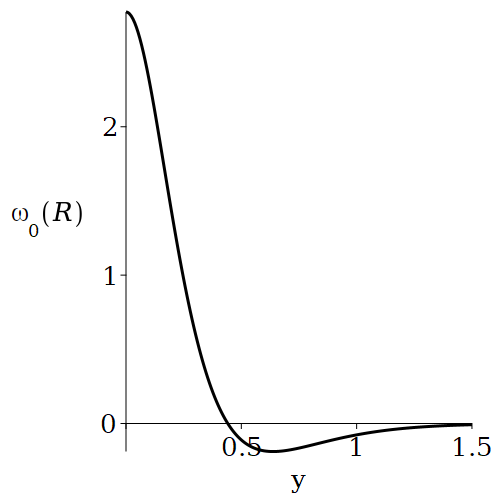}
\caption{\label{omega_0(R_nm)}The parameter of the exciton-phonon interaction $\omega
_{0}(R_{m^{\prime}m})$ (in arbitrary units) as a function of $y=R_{m^{\prime
}m}/a_{0}$ for $p_{e}\equiv m_{r}/m_{e}^{\ast}=2/3$ and $p_{h}\equiv
m_{r}/m_{h}^{\ast}=1/3$, where $1/m_{r}=1/m_{e}^{\ast}+1/m_{h}^{\ast}$ is the
inverse reduced mass.}
\end{center}
\end{figure}

In Eq.(\ref{eq.:dC_om(t)/dt3})
$\mathbf{R}_{m^{\prime}m}=\mathbf{R}_{m^{\prime}}-\mathbf{R}_{m}$,
$R_{m^{\prime}m}\equiv\left\vert \mathbf{R}_{m^{\prime}m}\right\vert $,
\begin{eqnarray}
\omega _{0}(\mathbf{R}_{nm}) &=&\omega _{0}(R_{nm})=\sum_{\mathbf{k}}\omega
_{0}(k)\exp (i\mathbf{k(R}_{n}-\mathbf{R}_{m}))  \nonumber \\
&=&\frac{L^{2}}{2\pi }\int_{0}^{\infty }dkk\omega _{0}(k)J_{0}(kR_{nm})
\label{eq:omega_0(R_nm)}
\end{eqnarray}%
is the Fourier transform of $\omega_{0}(k)$, $L^{2}$ corresponds to the area
of the thin film, $J_{0}(z)$ is the Bessel function of the first kind,
$\hbar\bar{W}_{0}=E_{c}(\mathbf{q}_{0})-E_{v}(\mathbf{q}_{0})+E_{0}$, and $\sum_{m}
|C_{0m}|^{2}=1-\sum_{\mathbf{q}}|G(\mathbf{q}|t)|^{2}$. The spontaneous decay
term $-\frac{1}{2}\Gamma_{0}(\omega)$ is discussed in Appendix A. This term
can be neglected when the formation time of the considered state or soliton is
much shorter than the spontaneous decay time, $\sim1/\Gamma_{0}(\omega)$. As
confirmed by our subsequent numerical calculations, this condition is
satisfied; therefore, the term is omitted in the analysis below for simplicity.
The graph of the Wannier exciton-LO phonon interaction parameter in the
coordinate space, $\omega_{0}(R_{m^{\prime}m})$, is shown in
Fig.\ref{omega_0(R_nm)}.%

In the case of local phonons $\omega_{0}(k)=\omega_{0}=const$ (since $|D_{l}^{pol}(\mathbf{k};00)|^{2}$ is a
constant\cite{Fainberg_Osipov2024JCP}),
and the nonlocal nonlinear term on the right-hand side of
Eq.(\ref{eq.:dC_om(t)/dt3}) becomes local:
\begin{eqnarray}
2i\sum_{m^{\prime }}|C_{0m^{\prime }}|^{2}\omega _{0}(\mathbf{R}_{m^{\prime
}m}) &=&i2\omega _{0}\sum_{m^{\prime }}|C_{0m^{\prime }}|^{2}\sum_{\mathbf{k}%
}\exp (i\mathbf{k(R}_{m^{\prime }}-\mathbf{R}_{m}))  \nonumber \\
&=&i2N\omega _{0}|C_{0m}|^{2},  \label{eq:Sum_n}
\end{eqnarray}%
since $\sum_{\mathbf{k}}\exp(i\mathbf{k(R}_{m^{\prime}}-\mathbf{R}%
_{m}))=N\delta_{m^{\prime}m}$. This leads to the Davydov soliton
\cite{Davydov82,Gelin2015} in the case of one-dimensional molecular aggregates.

Eq.(\ref{eq.:dC_om(t)/dt3}) can be written for the quantity $\tilde{C}%
_{0m}=C_{0m}\exp(i\bar{W}_{0}t)$ in the Cartesian coordinates and in the
continuum approximation, when%
\begin{equation}%
\sum_{\mathbf{R}^{\prime}}
\rightarrow\frac{l}{l_{0}^{3}}\int d^{2}R^{\prime}, \label{eq:sum(R')}%
\end{equation}
as follows:
\begin{widetext}
\begin{equation}
i\frac{\partial }{\partial t}\tilde{C}_{0}(x,y,t)+\Bigg[\frac{\hbar }{%
2M^{\ast }}\left( \frac{\partial ^{2}}{\partial x^{2}}+\frac{\partial ^{2}}{%
\partial y^{2}}\right) +\frac{2l}{l_{0}^{3}}\int \int dx^{\prime }dy^{\prime
}|\tilde{C}_{0}(x^{\prime },y^{\prime },t)|^{2}\omega _{0}(\sqrt{(x^{\prime
}-x)^{2}+(y^{\prime }-y)^{2}})\Bigg]\tilde{C}_{0}(x,y,t)=0
\label{eq.:dC_o(x,y,t)/dt}
\end{equation}%
\end{widetext}
where $l_{0}^{3}$ is the volume of the unit cell related to $l$, the thickness
of the thin film, and $L$ by $Nl_{0}^{3}=lL^{2}$. Equation
(\ref{eq.:dC_o(x,y,t)/dt}) is similar to the single nonlocal NLS equation (3)
of Ref.\cite{Malomed2022}.

\section{The solution of the continuum-approximation equation}

\label{Section:solution1}

\subsection{Analytical results}

The model (\ref{eq.:dC_o(x,y,t)/dt}) permits plane wave solutions of the form
\cite{Krolikowsk2004}%
\begin{equation}
\tilde{C}_{0}(\mathbf{R},t)=\sqrt{\rho_{0}}\exp(i\mathbf{k}_{0}\cdot
\mathbf{R}-i\beta t) \label{eq:C_0(R,t)}%
\end{equation}
where $\rho_{0}$ $>0$ is the wave intensity. As a special case, this equation
also includes $\tilde{C}_{0}(\mathbf{R},t)=\sqrt{\rho_{0}}=const$ when
$\mathbf{k}_{0}=\beta=0$ that corresponds to the superradiant state. One can
easily see that $\int d^{2}R\left\vert \tilde{C}_{0}(\mathbf{R}%
,t)\right\vert ^{2}=\rho _{0}L^{2}$. Bearing in mind the normalization, $%
\sum_{m}|C_{0m}|^{2}=1-\sum_{\mathbf{q}}|G(\mathbf{q}|t)|^{2}$, and Eq.(\ref%
{eq:sum(R')}), we get%
\begin{equation}
\rho _{0}=\frac{1}{N}\left[ 1-\sum_{\mathbf{q}}|G(\mathbf{q}|t)|^{2}\right]
\label{eq:rho_0}
\end{equation}

Substituting Eq.(\ref{eq:C_0(R,t)}) in Eq.(\ref{eq.:dC_o(x,y,t)/dt}), we
obtain the following dispersion relation:%
\begin{equation}
\beta -\frac{\hbar k_{0}^{2}}{2M^{\ast }}+\frac{2l}{l_{0}^{3}}\rho _{0}\int
\int \omega _{0}(\sqrt{(x^{\prime }-x)^{2}+(y^{\prime }-y)^{2}})dx^{\prime
}dy^{\prime }=0  \label{eq:dispersion1}
\end{equation}%
The last term on the left-hand side corresponds to the nonlocal interaction
and can be shown to vanish in the case of Wannier excitons interacting with LO
phonons via the Frohlich interaction mechanism. This simplification reflects
the fact that the exciton-phonon coupling is delocalized and isotropic in
nature. Therefore, Eq.(\ref{eq:dispersion1}) simplifies to the canonical
dispersion relation:
\begin{equation}
\beta=\frac{\hbar}{2M^{\ast}}k_{0}^{2} \label{eq:dispersion2}%
\end{equation}

Next, carry out the linear stability analysis of the plane wave solutions,
Eq.(\ref{eq:C_0(R,t)}). To this end, we assume that%
\begin{equation}
\tilde{C}_{0}(\mathbf{R},t)=[\sqrt{\rho_{0}}+a_{1}(\vec{\xi},t)]\exp
(i\mathbf{k}_{0}\cdot\mathbf{R}-i\beta t) \label{eq:C_0(R,t)pert}%
\end{equation}
where%
\begin{equation}
a_{1}(\vec{\xi},t)=\int\tilde{a}_{1}(\mathbf{k},t)\exp(i\mathbf{k}\cdot
\vec{\xi})d\mathbf{k}\quad\vec{\xi}=\mathbf{R}%
-\frac{d\beta}{dk_{0}}t=\mathbf{R}-\frac{\hbar}{M^{\ast}}k_{0}t
\label{eq:a_1(ksi,t)}%
\end{equation}
is the complex amplitude of the small perturbation referred to a coordinate
frame moving with the group velocity $\frac{\hbar}{M^{\ast}%
}k_{0}$.

Inserting Eq.(\ref{eq:C_0(R,t)pert}) into the nonlocal NLS equation
(\ref{eq.:dC_o(x,y,t)/dt}) and linearizing around the solution
(\ref{eq:C_0(R,t)}) yields the following evolution equation for small
perturbations:%
\begin{equation}
i\frac{\partial\bar{a}_{1}(\bar{\xi},t)}{\partial t}+\frac{1}{2}\nabla
_{\bar{\xi}}^{2}\bar{a}_{1}(\bar{\xi},t)+\frac{4l}{l_{0}^{3}}\rho_{0}%
\int\operatorname{Re}[\bar{a}_{1}(\bar{\xi}^{\prime},t)]\omega_{0}(\bar{\xi
}^{\prime}-\bar{\xi})d\bar{\xi}^{\prime}=0 \label{eq:a_1(ksi,t)2}%
\end{equation}
where we moved to new coordinates $\bar{R}=\sqrt{\frac{M^{\ast}}{\hbar}%
}\mathbf{R}$, and correspondingly $\bar{\xi}=\sqrt{\frac{M^{\ast}}{\hbar}}%
\vec{\xi}$ , with $a_{1}(\vec{\xi},t)$ replaced by $\bar{a}_{1}(\bar{\xi},t)$.
All newly introduced variables, including the dimensionless quantities,
are summarized in Appendix F.

Decomposing the perturbation into real and imaginary parts, $\bar{a}%
_{1}(\mathbf{\bar{\xi}},t)=u(\mathbf{\bar{\xi}},t)+iv(\mathbf{\bar{\xi}},t)$,
we obtain two coupled equations
\begin{equation}
\frac{\partial u(\mathbf{\bar{\xi}},t)}{\partial t}+\frac{1}{2}\nabla
_{\bar{\xi}}^{2}v(\mathbf{\bar{\xi}},t)=0 \label{eq:u}%
\end{equation}%
\begin{equation}
\frac{\partial v(\mathbf{\bar{\xi}},t)}{\partial t}-\frac{1}{2}\nabla
_{\bar{\xi}}^{2}u(\mathbf{\bar{\xi}},t)-\frac{4l}{l_{0}^{3}}\rho_{0}\int
u(\mathbf{\bar{\xi}}^{\prime},t)\omega_{0}(\bar{\xi}^{\prime}-\bar{\xi}%
)d\bar{\xi}^{\prime}=0 \label{eq:v}%
\end{equation}
By introducing the Fourier transforms, we get%
\begin{eqnarray}
u(\bar{\xi},t) &=&\frac{1}{\sqrt{N}}\sum_{\mathbf{k}}u(\bar{k},t)\exp (i%
\mathbf{\bar{k}\bar{\xi}}),  \nonumber \\
v(\bar{\xi},t) &=&\frac{1}{\sqrt{N}}\sum_{\mathbf{\bar{k}}}v(\bar{k},t)\exp
(i\mathbf{\bar{k}\bar{\xi}}),
\end{eqnarray}%
where in the continuum approach,
\begin{eqnarray}
u(\bar{k},t) &=&\frac{1}{\sqrt{N}}\frac{l}{l_{0}^{3}}\int d^{2}\bar{\xi}u(%
\mathbf{\bar{\xi}},t)\exp (-i\mathbf{\bar{k}\bar{\xi}}),\quad   \nonumber \\
v(\bar{k},t) &=&\frac{1}{\sqrt{N}}\frac{l}{l_{0}^{3}}\int d^{2}\bar{\xi}v(%
\mathbf{\bar{\xi}},t)\exp (-i\mathbf{\bar{k}\bar{\xi}}),
\end{eqnarray}%
$\bar{k}=\sqrt{\frac{\hbar }{M^{\ast }}}k$. Thus, our linearized system is reduced to a set of ordinary differential
equations for Fourier transforms $u(\bar{k},t)$ and $v(\bar{k},t)$, which can
be written in the matrix form as%

\begin{equation}
\frac{\partial}{\partial t}\mathbf{Z}=\hat{K}\mathbf{Z} \label{eq:Z}%
\end{equation}
where vector $\mathbf{Z}$ and matrix $\hat{K}$ are
\begin{equation}
\mathbf{Z}=\left(
\begin{array}
[c]{c}%
u(\bar{k},t)\\
v(\bar{k},t)
\end{array}
\right)  ,\quad\hat{K}=\left(
\begin{array}
[c]{cc}%
0 & \frac{1}{2}\bar{k}^{2}\\
-\frac{1}{2}\bar{k}^{2}+4N\rho_{0}\omega_{0}(-\bar{k}) & 0
\end{array}
\right)  . \label{eq:Z,K}%
\end{equation}
The eigenvalues $\lambda$ of the matrix $\hat{K}$ are defined by%
\begin{equation}
\lambda^{2}=-\frac{\bar{k}^{4}}{4}\left(  1-\frac{\rho_{0}}{\rho_{cr}}\right)
\label{eq:lambda^2}%
\end{equation}
where
\begin{equation}
\rho_{cr}(\bar{k})=\frac{\bar{k}^{2}}{8N\omega_{0}(-\bar{k})}
\label{eq:rho_cr}%
\end{equation}
denotes a critical value of the square of amplitude $\rho_{0}$. Hence we
conclude that $\lambda^{2}<0$ (stability holds) when $\rho_{0}<\rho_{cr}$.

For small $\bar{k}$, $\omega_{0}(-\bar{k})\sim\bar{k}^{2}$, and the critical
value $\rho_{cr}$ does not depend on wave number. Indeed, using
Eq.(\ref{eq:omega_0(k)}) above, Eq.(12) of Ref.\cite{Fainberg_Osipov2024JCP}
and Eq.(\ref{eq:rho_cr}), we get%

\begin{equation}
\rho_{cr}=\frac{32}{9\pi\alpha_{e}(p_{h}^{2}-p_{e}^{2})^{2}}\left(
\frac{l_{0}}{a_{0}}\right)  ^{3}\frac{1}{a_{0}}\sqrt{\frac{2p_{h}\hbar}%
{\omega_{l}M^{\ast}}} \label{eq:(rho_0)_cr}%
\end{equation}
where $\alpha_{e}=\frac{e^{2}}{2\hbar}\left(  \frac{1}{\varepsilon_{\infty}%
}-\frac{1}{\varepsilon_{0}}\right)  \sqrt{\frac{2m_{e}^{\ast}}{\hbar\omega
_{l}}}$ is the Frohlich coupling constant
\cite{Frohlich54,Feynman98,Jai_Singh94,Miyata17}, $e$ is the charge of
carrier, $\varepsilon_{\infty}$ and $\varepsilon_{0}$ are optical and static
dielectric constants, $p_{i}\equiv m_{r}/m_{i}^{\ast}$, $i=e,h$;
$1/m_{r}=1/m_{e}^{\ast}+1/m_{h}^{\ast}$ is the inverse reduced mass. Small
amplitude plane waves with $\rho_{0}\leq\rho_{cr}$ are modulationally stable,
whereas modulational instability appears at high amplitudes, $\rho_{0}%
>\rho_{cr}$. This instability is long-wave \cite{Malomed2025}, bearing in mind
the $\bar{k}^{2}$-expansion of $\omega_{0}(-\bar{k})$.

Using characteristic values of the parameters: $\sqrt{\frac{\hbar}{2\omega
_{l}M^{\ast}}}\approx\frac{a_{0}}{2}$, $p_{h}=0.4,p_{e}=0.6$
\cite{Fainberg_Osipov2024JCP}, $a_{0}\simeq55\mathring{A}$, $l=2a_{0}$,
$l_{0}=6\mathring{A}$ for methylammonium lead iodide perovskite \cite{Frohna18}
and $\alpha_{e}\simeq2.5$ in hybrid organic-inorganic perovskites \cite{Silva2020JPCL}, we obtain $\rho_{cr}\simeq0.01$.
To estimate the
exciton radius $a_{0}$, we used the expression reported in Ref. \cite%
{Fainberg_Osipov2024JCP}: $a_{0}=\hbar \varepsilon _{0}/(\alpha
cp_{e}p_{h}M^{\ast })$, where $\alpha =1/137$ is the fine-structure
constant. In the numerical estimate, we set $\varepsilon _{0}=10$ and $%
M^{\ast }=0.4m_{e}=3.\allowbreak 64\cdot 10^{\mathbf{-}28}g$, following Ref.%
\cite{Fainberg_Osipov2024JCP}.

It should be noted that though the plane wave solution does not correspond to
the circular symmetry of the problem, its special case with $\beta=k_{0}=0$
that matches the superradiant state, does have the circular symmetry.
Therefore, Eq.(\ref{eq:(rho_0)_cr}) is of direct interest for the evaluation
of the superradiant state stability in the quasi-2D structures possessing the
circular symmetry.

\subsection{Numerical results}
The numeric solution of Eq.(\ref{eq.:dC_o(x,y,t)/dt}) (see dimensionless
Eq.(\ref{eq.:dC_o(x,y,t)/dt}) in Appendix C) was obtained using approximation,
Eq.(\ref{eq:omega_0(R)appr}) of Appendix B, for the parameter of
exciton-phonon interaction $\omega_{0}(R_{m^{\prime}m})$. The initial
condition was taken as a plane wave with a noise-perturbed amplitude, $A(%
\tilde{x},\tilde{y},0)=\sqrt{\rho _{0}}+0.01\cdot noise$, where $noise=%
rand(\tilde{x},\tilde{y})$. The computational domain was defined by $\tilde{%
x}_{\min }=-60,$ $\tilde{x}_{\max }=64$, $\tilde{y}_{\min }=-64,$ and $%
\tilde{y}_{\max }=60$ with $NUM_{X}=NUM_{Y}=61$ (grid size). The spatial
step was $d\tilde{x}=d\tilde{y}=2.0(6)$, and the time step was $d\tilde{t}%
=0.01$. Zero Dirichlet boundary conditions were imposed. Time integration
was performed using the fourth-order Runge-Kutta method.

\subsubsection{Subcritical case}

The numerical results obtained for a spatially homogeneous initial condition
with noise for subcritical case ($\rho_{0}=0.5\rho_{cr,0}$ where $\rho
_{cr,0}\equiv\rho_{cr}(\bar{k}\rightarrow0)\simeq0.01$) are shown in
Fig.\ref{fig:subcritical(c,d,e)} in dimensionless coordinates. Dimensional
values {}{}are obtained by multiplying by the characteristic length
$D=\sqrt[4]{\hbar l_{0}^{3}/\left(  4G_{0}M^{\ast}l\right)  }$ and
characteristic time $\tau=\sqrt{M^{\ast}l_{0}^{3}/\left(  G_{0}l\hbar\right)
}$, respectively. For the characteristic values {}{}of parameters,
characteristic length and time are equal to $D\approx8.9\cdot10^{\mathbf{-}%
8}cm$ and $\tau\approx5.48\cdot10^{\mathbf{-}15}s$, respectively.
Fig.\ref{fig:subcritical(c,d,e)} illustrates good stability of the initial
state for the subcritical case.%

\begin{figure*}
\begin{center}
\includegraphics[scale=0.33]
{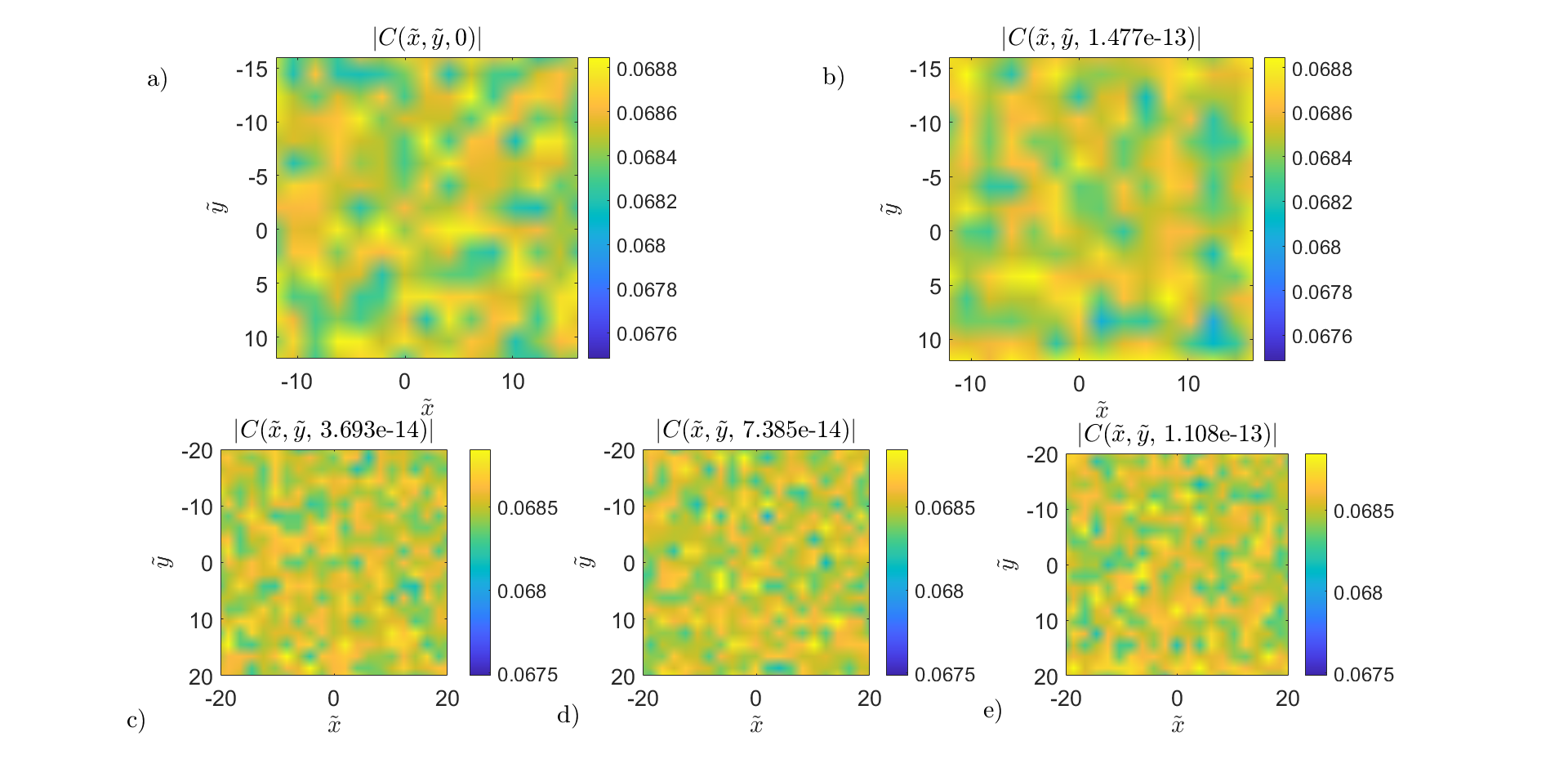}%
\caption{\label{fig:subcritical(c,d,e)} Modulus of $\tilde{C}_{0}(\tilde{x},\tilde{y},t)$ at $t=0$,
$\tilde{C}_{0}(\tilde{x},\tilde{y},0)=\sqrt{\rho_{0}}\left[  1+0.01\cdot
noise\left(  \tilde{x},\tilde{y}\right)  \right]  $, (a); $t=0.27%
\protect\tau $ (b), $t=6.73\protect\tau $ (c), $t=13.5\protect\tau $ (d) and
$t=20.26\protect\tau $ (e) for
subcritical case ($\rho_{0}=0.5\rho_{cr,0}$) in dimensionless coordinates. }
\end{center}
\end{figure*}

\subsubsection{Instability analysis}

\begin{figure*}
\begin{center}
\includegraphics[scale=0.33]{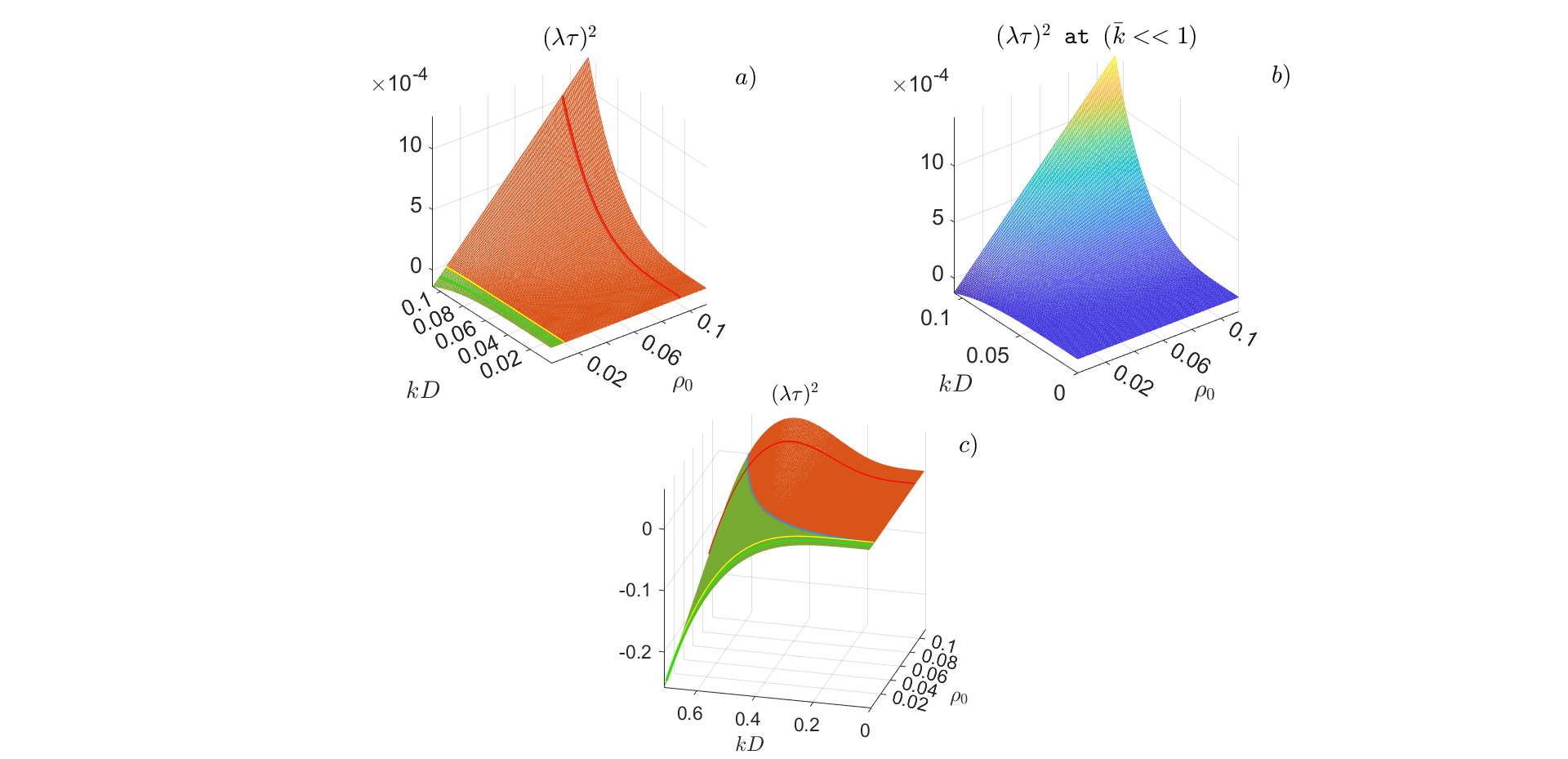}%
\caption{Square of dimensionless increment $(\lambda\tau)^{2}$,
Eq.(\ref{eq:lambda^2}), as a function of wave intensity $\rho_{0}$ and $kD$.
Calculations using exact formula Eq.(\ref{eq:rho_cr}) (a,c), and the long-wave
limit (small $k$) Eq.(\ref{eq:(rho_0)_cr}) (b). The green line is drawn at
the level $\rho_{0}=0.5\rho_{cr,0}$, the yellow line - at the level $\rho
_{0}=\rho_{cr,0}$, and the red line - at the level $\rho_{0}=10\rho_{cr,0}$
(a,c). The blue line is drawn at the level $(\lambda\tau)^{2}=0$ (c).}%
\label{fig:increment}%
\end{center}
\end{figure*}

Fig.\ref{fig:increment} shows the square of the dimensionless increment
$(\lambda\tau)^{2}$ as a function of the wave intensity $\rho_{0}$ and
dimensionless wave number $kD$. This allows us to find the value of $kD$ for
which the increment is maximum for some value of $\rho_{0}>\rho_{cr,0}$. For
$\rho_{0}=10\rho_{cr,0}\simeq0.1$, the maximum increment value is achieved at
$kD\approx0.43$. In this case, the value of the increment is equal to
$\lambda\tau\approx0.22$ that corresponds to the characteristic time of
developing the instability $\tau_{0}\approx4.53\tau=2.48\cdot10^{\mathbf{-}%
14}s$.%

\subsubsection{Modeling the development of instability}

The numerical results obtained for a spatially homogeneous initial condition
with noise for a supercritical case ($\rho_{0}=10\rho_{cr,0}\simeq0.1$) are
shown in Fig.\ref{fig:supercritical(c,d,e)} in dimensionless coordinates.
Fig.\ref{fig:supercritical(c,d,e)} illustrates the development of instability
of a plane wave in the supercritical case. A plane wave with a noise perturbed
amplitude was used as the initial condition. It is evident that this
perturbation grows as the wave propagates.%

\begin{figure*}
\begin{center}
\includegraphics[scale=0.33]{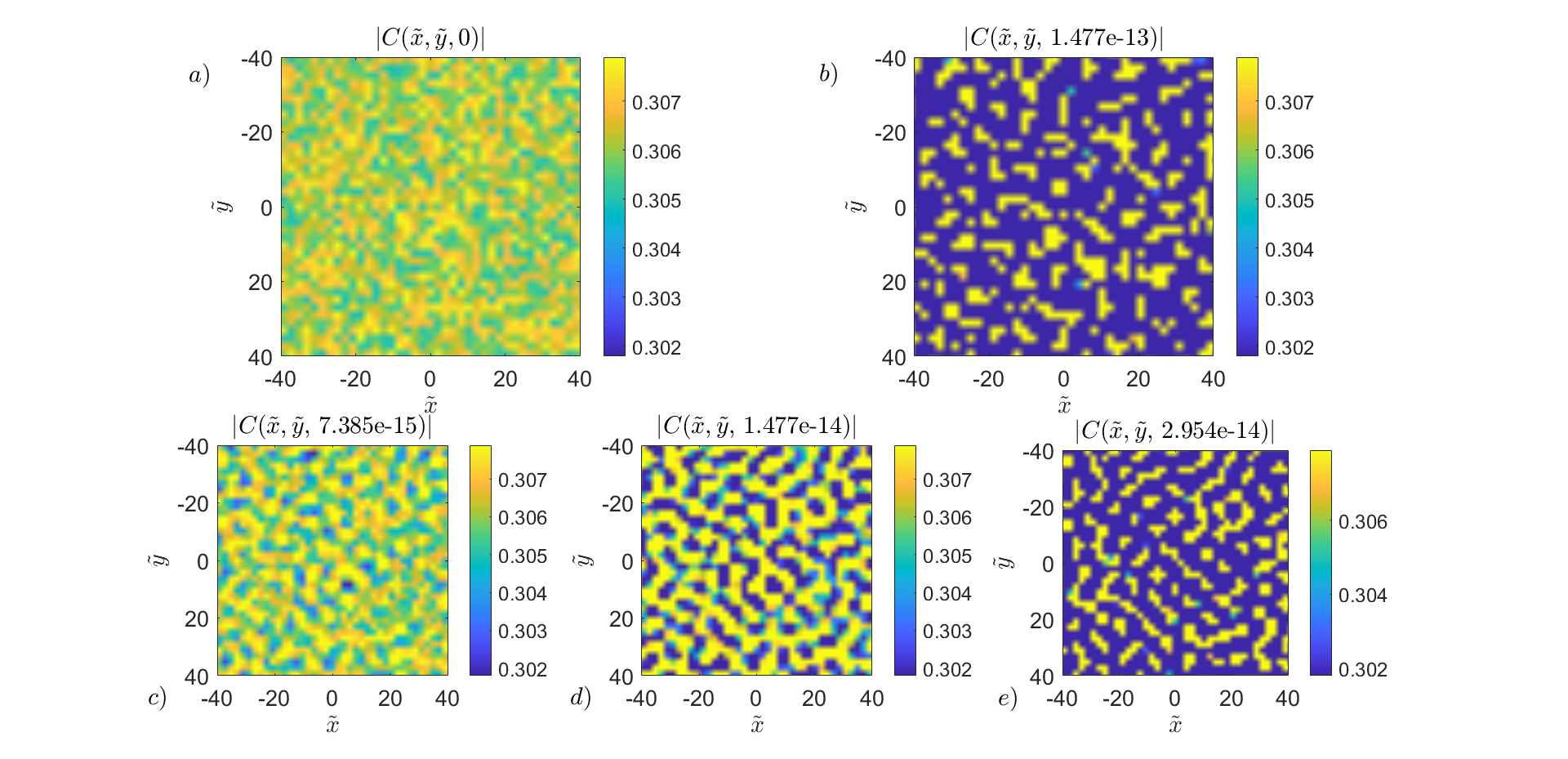}%
\caption{Modulus of $\tilde{C}_{0}(\tilde{x},\tilde{y},t)$ at $t=0$,
$\tilde{C}_{0}(\tilde{x},\tilde{y},0)=\sqrt{\rho_{0}}\left[  1+0.01\cdot
noise\left(  \tilde{x},\tilde{y}\right)  \right]  $, (a); $t=1.48\cdot
10^{\mathbf{-}13}s$ (b), $t=\tau_{0}/2$ (c), $t=\tau_{0}$ (d) and $t=2\tau
_{0}$ (e) for supercritical case ($\rho_{0}=10\rho_{cr}$) in dimensionless
coordinates. }%
\label{fig:supercritical(c,d,e)}%
\end{center}
\end{figure*}

\section{The limit of the weakly nonlocal NLS equation}

\label{Section:weakly nonlocal}

\subsection{The derivation of the weakly nonlocal NLS equation}

The solution of Eq.(\ref{eq.:dC_o(x,y,t)/dt}) depends on the ratio of the
widths of the functions $|\tilde{C}_{0}(x^{\prime},y^{\prime},t)|^{2}$ and
$\omega_{0}(\mathbf{r}^{\prime}-\mathbf{r})$ \cite{Krolikowsk2004}. According
to Fig.\ref{omega_0(R_nm)}, one has $\omega_{0}(\mathbf{r}^{\prime}%
-\mathbf{r})>0$ ("focusing nonlinearity") at $0\leq|\mathbf{r}^{\prime
}-\mathbf{r}|\leq0.45a_{0}$, and $\omega_{0}(\mathbf{r}^{\prime}%
-\mathbf{r})<0$ ("defocusing nonlinearity") at $0.45a_{0}\leq|\mathbf{r}%
^{\prime}-\mathbf{r}|\leq1.5a_{0}$ where the strength of the "defocusing
nonlinearity" is much smaller than that of the "focusing nonlinearity". In the
present case, we have $L>>a_{0}$. If initially the system is in the
superradiant state $|SR\rangle=\frac{1}{\sqrt{N}}%
{\displaystyle\sum\limits_{m=1}^{N}}
|me\rangle$, then the width of $|\tilde{C}_{0}(x^{\prime},y^{\prime}%
,t=0)|^{2}$ is $\sim L$, which is much larger than that of $\omega
_{0}(\mathbf{r}^{\prime}-\mathbf{r})$. This is the case of the weak
nonlocality limit \cite{Krolikowsk2004}. In this case one can expand
$|\tilde{C}_{0}(\mathbf{R},t)|^{2}$ in the Taylor series and retain only the
first significant terms. Then the integral on the right-hand side of
Eq.(\ref{eq.:dC_o(x,y,t)/dt}) takes the form
\begin{widetext}
\begin{equation}
\int_{-\infty }^{\infty }\int_{-\infty }^{\infty }|\tilde{C}_{0}(x^{\prime
},y^{\prime },t)|^{2}\omega _{0}(\sqrt{\breve{x}^{2}+\breve{y}^{2}}%
)dx^{\prime }dy^{\prime }=\frac{\pi }{2}\left[ \frac{\partial ^{2}|\tilde{C}%
_{0}(x,y,t)|^{2}}{\partial x^{2}}+\frac{\partial ^{2}|\tilde{C}%
_{0}(x,y,t)|^{2}}{\partial y^{2}}\right] \int_{0}^{\infty }drr^{3}\omega
_{0}(r)  \label{eq:nonlocal_term}
\end{equation}%
\end{widetext}
where $\breve{x}=x^{\prime}-x,\breve{y}=y^{\prime}-y$ and $\breve{x}%
^{2}+\breve{y}^{2}=r^{2}$. Here the zero-constant part of the matrix element
$D_{l}^{pol}(\mathbf{k};00)$ cancels the local nonlinear term.
Indeed, in the local approximation the last term on the
right-hand-side of Eq.(\ref{eq.:dC_om(t)/dt3}) becomes%
\begin{eqnarray}
&&i2\tilde{C}_{0}(\mathbf{R},t)\sum_{\mathbf{R}^{\prime }}|\tilde{C}_{0}(%
\mathbf{R}^{\prime },t)|^{2}\omega _{0}(\mathbf{R}^{\prime }-\mathbf{R})
\notag \\
&\approx &i2\tilde{C}_{0}(\mathbf{R},t)|\tilde{C}_{0}(\mathbf{R}%
,t)|^{2}\sum_{\mathbf{R}^{\prime }}\omega _{0}(\mathbf{R}^{\prime }-%
\mathbf{R})  \label{eq:local_limit2}
\end{eqnarray}%
Using the Fourier transform of $\omega _{0}(\mathbf{R}^{\prime }-\mathbf{R})$,%
Eq.(\ref{eq:omega_0(R_nm)}), we get
\begin{equation*}
\sum_{\mathbf{R}^{\prime }}\omega _{0}(\mathbf{R}^{\prime }-\mathbf{R%
})=\sum_{\mathbf{k}}\omega _{0}(k)\exp (-i\mathbf{kR})\sum_{\mathbf{R%
}^{\prime }}\exp (i\mathbf{k}(\mathbf{R}^{\prime })
\end{equation*}
on the right-hand-side of Eq.(\ref{eq:local_limit2}). The sum is equal to $%
\sum_{\mathbf{R}^{\prime }}\exp (i\mathbf{kR}^{\prime })=N\delta _{%
\mathbf{k}0}$ that gives
\begin{equation*}
\sum_{\mathbf{R}^{\prime }}\omega _{0}(\mathbf{R}^{\prime }-\mathbf{R%
})=N|\tilde{C}_{0}(\mathbf{R},t)|^{2}\omega _{0}(k=0)=0,
\end{equation*}
since $\omega _{0}(k=0)=0$ for
the Wannier exciton coupled to phonons via the LO phonon-exciton
Frohlich interaction \cite{Fainberg_Osipov2024JCP}.

\begin{figure}
[ptb]
\begin{center}
\includegraphics[scale=0.5]{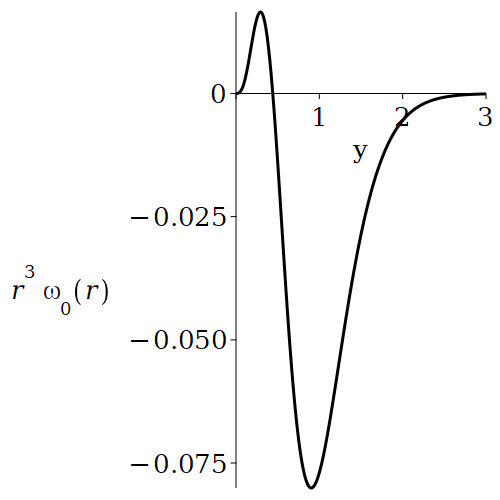}%
\caption{Function $r^{3}\omega_{0}(r)$ (in arbitrary units) as a function of
$y=r/a_{0}$ for $p_{e}=2/3$ and $p_{h}=1/3$.}%
\label{fig:r^3xomega_0(r)}%
\end{center}
\end{figure}

The integrand of $\int_{0}^{\infty}drr^{3}\omega_{0}(r)$ is shown in
Fig.\ref{fig:r^3xomega_0(r)}, which implies that

$\int_{0}^{\infty}drr^{3}\omega_{0}(r)<0$. Using transformations $\bar
{x}=\sqrt{\frac{M^{\ast}}{\hbar}}x$ and $\bar{y}=\sqrt{\frac{M^{\ast}}{\hbar}%
}y$, we finally get%
\begin{equation}
i\frac{\partial}{\partial t}\bar{C}_{0}(\bar{x},\bar{y},t)+\frac{1}{2}%
\nabla^{2}\bar{C}_{0}(\bar{x},\bar{y},t)=\bar{J}\bar{C}_{0}(\bar{x},\bar{y},t)\nabla^{2}|\bar{C}%
_{0}(\bar{x},\bar{y},t)|^{2}
\label{eq.:dC_o(x,y,t)/dt3}%
\end{equation}
where $\nabla^{2}=\frac{\partial^{2}}{\partial\bar{x}^{2}}+\frac{\partial^{2}%
}{\partial\bar{y}^{2}}$, $\bar{J}=J\frac{M^{\ast}}{\hbar}$, and the integral
is
\begin{equation}
J\equiv\frac{\pi l}{l_{0}^{3}}\left\vert \int_{0}^{\infty}drr^{3}\omega
_{0}(r)\right\vert >0 \label{eq.:J}%
\end{equation}

\subsection{Solution of Eq.(\ref{eq.:dC_o(x,y,t)/dt3})}

Obviously, the coherent state $\bar{C}_{0}(\bar{x},\bar{y},t)=const$ is the
solution of Eq.(\ref{eq.:dC_o(x,y,t)/dt3}). In one-dimensional case
Eq.(\ref{eq.:dC_o(x,y,t)/dt3}) is reduced to Eq.(B1) of
Ref.\cite{Porkolab1976}. The solutions to the 1D version of
Eq.(\ref{eq.:dC_o(x,y,t)/dt3}) are cusped dark and bright solitons
\cite{Porkolab1976}.

Next, we perform the stability analysis for the plane wave solution of
Eq.(\ref{eq.:dC_o(x,y,t)/dt3}), $\bar{C}_{0}(\mathbf{\bar{R}},t)=\sqrt
{\rho_{0}}\exp[i(\mathbf{\bar{k}}_{0}\cdot\mathbf{\bar{R}}-\beta t)]$, where
$\beta=\frac{1}{2}\bar{k}_{0}^{2}$ (see Eq.(\ref{eq:dispersion2})). We carry
out the linear stability analysis of the plane wave solutions similar to that
presented in Section \ref{Section:solution1}. We assume that $\bar{C}%
_{0}(\mathbf{\bar{R}},t)=[\sqrt{\rho_{0}}+\bar{b}_{1}(\mathbf{\bar{\xi}%
},t)]\exp(i\mathbf{\bar{k}}_{0}\cdot\mathbf{\bar{R}}-i\beta t)$. Inserting
this into Eq.(\ref{eq.:dC_o(x,y,t)/dt3}) and linearizing yields the evolution
equation for the perturbation $\bar{b}_{1}(\mathbf{\bar{\xi}},t)$. Decomposing
the perturbation into the real and imaginary parts, $\bar{b}_{1}%
(\mathbf{\bar{\xi}},t)=u_{b}(\mathbf{\bar{\xi}},t)+iv_{b}(\mathbf{\bar{\xi}%
},t)$, we obtain two coupled equations%
\begin{equation}
\frac{\partial u_{b}(\mathbf{\bar{\xi}},t)}{\partial t}+\frac{1}{2}%
\nabla_{\bar{\xi}}^{2}v_{b}(\mathbf{\bar{\xi}},t)=0 \label{eq:u_b}%
\end{equation}%
\begin{equation}
\frac{\partial v_{b}(\mathbf{\bar{\xi}},t)}{\partial t}-\frac{1}{2}(1-4\bar
{J}\rho_{0})\nabla_{\bar{\xi}}^{2}u_{b}(\mathbf{\bar{\xi}},t)=0 \label{eq:v_b}%
\end{equation}
By introducing the Fourier transforms of $u_{b}(\mathbf{\bar{\xi}},t)$ and
$v_{b}(\mathbf{\bar{\xi}},t)$, $u_{b}(\bar{k},t)$ and $v_{b}(\bar{k},t)$,
respectively, we get%

\begin{equation}
\frac{\partial}{\partial t}u_{b}(\bar{k},t)=\frac{1}{2}v_{b}(\bar{k},t)\bar
{k}^{2} \label{eq:u_b(k)}%
\end{equation}

\begin{equation}
\frac{\partial}{\partial t}v_{b}(\bar{k},t)=-\frac{1}{2}\bar{k}^{2}%
(1-4\rho_{0}\bar{J})u_{b}(\bar{k},t) \label{eq:v_b(k)}%
\end{equation}
The eigenvalues $\lambda$ of the matrix similar to matrix $\hat{K}$, see
Eqs.(\ref{eq:Z,K}), for Eqs.(\ref{eq:u_b(k)}) and (\ref{eq:v_b(k)}) are found
to be%
\begin{equation}
\lambda^{2}=-\bar{k}^{4}\rho_{0}\left(  \frac{1}{4\rho_{0}}-\bar{J}\right)
\label{eq:lambda^2_2}%
\end{equation}
Thus we get $\lambda^{2}<0$ (stability) when $1/(4\rho_{0})>\bar{J}$. In other
words, a critical value of the intensity is
\begin{equation}
\rho_{cr,2}=\frac{1}{4\bar{J}} \label{eq:(rho_0)_cr2}%
\end{equation}
It is easy to show that both values of the intensity critical value defined by
Eqs.(\ref{eq:(rho_0)_cr}) and (\ref{eq:(rho_0)_cr2}) are almost identical with
good accuracy for the present parameter values.

It is worthy to note that the sign of the nonlocal nonlinearity in our case is
the same as that in the two-fluid theory of so-called cusp 1D solitons
\cite{Porkolab1976}.

\section{The interaction with acoustic phonons}

\label{Section:acoustic}

Above we considered the dominant exciton-phonon interaction in the hybrid
perovskites, viz., the LO phonon-exciton Frohlich interaction, in which case
the local nonlinear term is absent. However, there are indications that one
should move beyond the Fr\"{o}hlich picture and consider the interplay between
the short-range (exciton-acoustic phonon) and long-range (exciton-LO phonon)
lattice coupling in the systems under the consideration \cite{Silva2020JPCL},
since short-range interactions play an important role in
determining the spatial extent of the polaronic wave function in hybrid
organic-inorganic perovskites. According to Ref.\cite{Silva2020JPCL}, such
short-range interactions may arise from the dynamical lattice fluctuations induced
by the relative motion of the organic and inorganic sublattices, as well as
from the disorder intrinsic to the ionic lattice. While phonons within the
lattice plane interact via the long-range Frohlich coupling, lattice motion
across adjacent lattice planes manifests itself through the image charge as a
short-range potential, thereby contributing to the overall polaronic
coupling.
In that case we demonstrate below that a local nonlinear term will appear in
addition to the nonlocal one, and the consideration may be reduce to that of
Ref.\cite{Garcia_Malomed2003}.

\subsection{The case of only acoustic phonons}

For acoustic phonons, the parameter of exciton-phonon interaction
$\omega_{0,ac}(k)$ is defined by the same formula as that for LO phonons, see
Eqs.(\ref{eq:omega_0(k)}), with the only difference that the LO frequency
$\omega_{l}$ and the electronic matrix element for the coupling function of
the Wannier exciton-LO phonon interaction $D_{l}^{pol}(\mathbf{k};00)$ should
be replaced with the acoustic phonon frequency $\omega_{ac}(k)=\bar{u}k$ and
the electronic matrix element for the coupling function of the Wannier
exciton-acoustic phonon interaction $D_{ac}(\mathbf{k};00)$, respectively:
\begin{equation}
\omega_{0,ac}(k)=\frac{\omega_{ac}(k)}{\omega_{ac}^{2}(k)+\gamma^{2}}%
|D_{ac}(\mathbf{k};00)|^{2} \label{eq:omega_0ac(k)}%
\end{equation}
where $\bar{u}$ is an average sound velocity.

According to Ref.\cite{Toyozawa59},%
\begin{equation}
D_{ac}(\mathbf{k};nn^{\prime})=\sqrt{\frac{2}{9\hbar NM\bar{u}}}\sqrt{k}%
[C_{c}Q_{e}(\mathbf{k};nn^{\prime})-C_{h}Q_{h}(\mathbf{k};nn^{\prime})]
\label{eq.:D_ac}%
\end{equation}
where $Q_{e(h)}(\mathbf{k};nn^{\prime})\equiv\int d\mathbf{r}\psi_{n}^{\ast
}(\mathbf{r})\psi_{n^{\prime}}(\mathbf{r})\exp(\pm ip_{e(h)}\mathbf{kr})$ is
the Fourier transform of the charge distributions of the electron (hole) in
the internal motion \cite{Toyozawa58,Toyozawa59} representing the effective
distribution of the electron (hole) charge for a particular phonon
$\mathbf{k}$; $M$ denotes the mass of a unit cell; $-2C/3$ is the so called
deformation potential, that is, the energy change of the band bottom (or the
top) due to unit dilation of lattice. The quantities $Q_{e(h)}(\mathbf{k}%
;nn^{\prime})$ were calculated analytically for the quasi-2D model under
consideration and $n=n^{\prime}=0$ in Ref.\cite{Fainberg_Osipov2024JCP}%
\begin{equation}
Q_{e(h)}(\mathbf{k};00)=\left(  1+(\frac{p_{e(h)}ka_{0}}{4})^{2}\right)
^{\mathbf{-}3/2} \label{eq:Q_e,h}%
\end{equation}

Then the parameter of the exciton-phonon interaction for the acoustic phonons
$\omega_{0,ac}(k)$, Eq.(\ref{eq:omega_0ac(k)}), is
\begin{widetext}
\begin{equation}
\omega_{0,ac}(k)=\frac{2C_{h}^{2}}{9\hbar NM\bar{u}^{2}}\left[  \left(
1+(\frac{p_{h}ka_{0}}{4})^{2}\right)  ^{\mathbf{-}3/2}-\frac{C_{c}}{C_{h}%
}\left(  1+(\frac{p_{e}ka_{0}}{4})^{2}\right)  ^{\mathbf{-}3/2}\right]  ^{2}
\label{eq:omega_0,ac(k)}%
\end{equation}
\end{widetext}
at $\gamma\rightarrow0$. Given that the hole-lattice interaction is stronger
than electron-lattice interaction \cite{Jai_Singh94}, the effective mass of
holes exceeds that of electrons, resulting in $C_{h}>C_{c}$. The plot of
function $\omega_{0,ac}(k)$ is shown in Fig.\ref{fig:omega_0,ac(k)}.%

\begin{figure}
[ptb]
\begin{center}
\includegraphics[scale=0.5]{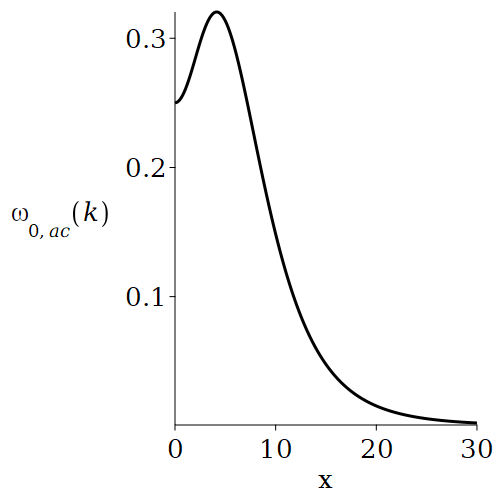}%
\caption{\label{fig:omega_0,ac(k)}Parameter of exciton-acoustic phonon interaction $\omega_{0,ac}(k)$
(in arbitrary units) as a function of $x=ka_{0}$ for $\gamma\rightarrow0$,
$p_{e}=2/3$, $p_{h}=1/3$, $C_{c}/C_{h}=1/2$. }
\end{center}
\end{figure}

\begin{figure}
[ptb]
\begin{center}
\includegraphics[scale=0.5]{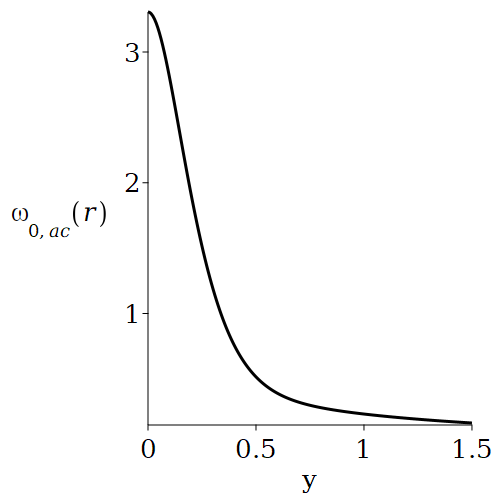}
\caption{\label{fig:omega_0ac(r)}Parameter of exciton-acoustic phonon interaction $\omega_{0,ac}(r)$
(in arbitrary units) as a function of $y=r/a_{0}$ for $p_{e}=2/3$, $p_{h}%
=1/3$, $C_{c}/C_{h}=1/2$. }
\end{center}
\end{figure}

Using Eq.(\ref{eq:omega_0(R_nm)}), one can calculate the Fourier-transform of
$\omega_{0,ac}(k)$, $\omega_{0,ac}(r)$ that is shown in
Fig.\ref{fig:omega_0ac(r)}.%

One can see that the width of the function $\omega_{0,ac}(r)$ is about the
exciton radius $a_{0}$ that is much smaller than that of $|\tilde{C}%
_{0}(R^{\prime},t=0)|^{2}$ when the system initially is in the superradiant
state (see Section \ref{Section:weakly nonlocal}). This is the case of the
weak nonlocality limit \cite{Krolikowsk2004} when formally one can expand
$|\tilde{C}_{0}(R,t)|^{2}$ in the Taylor series and retain only the first
significant term. Then the nonlocal nonlinear term on the right-hand side of
Eq.(\ref{eq.:dC_om(t)/dt3}) becomes local for the interaction with acoustic
phonons in contrast to the purely nonlocal term for the interaction with the
LO phonons. Indeed, one obtains%
\begin{eqnarray}
&&\sum_{\mathbf{R}^{\prime }}|\tilde{C}_{0}(\mathbf{R}^{\prime
},t)|^{2}\omega _{0,ac}(\mathbf{R}^{\prime }-\mathbf{R})  \notag \\
&\approx &|\tilde{C}_{0}(\mathbf{R},t)|^{2}\sum_{\mathbf{R}^{\prime }}\omega
_{0,ac}(\mathbf{R}^{\prime }-\mathbf{R})  \notag \\
&=&|\tilde{C}_{0}(\mathbf{R},t)|^{2}\sum_{\mathbf{k}}\omega _{0,ac}(k)\sum_{%
\mathbf{R}^{\prime }}\exp [i\mathbf{k}(\mathbf{R}^{\prime }-\mathbf{R})]
\notag \\
&=&N|\tilde{C}_{0}(\mathbf{R},t)|^{2}\omega _{0,ac}(k=0)\neq 0
\label{eq:local_limit}
\end{eqnarray}%
as $\sum_{\mathbf{R}^{\prime}}
\exp(i\mathbf{kR}^{\prime})=N\delta_{\mathbf{k}0}$. This term is not zero,
since $\omega_{0,ac}(k=0)\ne 0$ in contrast to $\omega_{0}(k=0)=0$ for the
interaction with the LO phonons.

\subsection{The interaction with both the LO and acoustic phonons}

For the interaction with both the LO \ and acoustic phonons,
Eq.(\ref{eq.:dC_om(t)/dt3}) can be written as%

\begin{widetext}
\begin{equation}
i\frac{d}{dt}\tilde{C}_{0m}(t)+[\frac{\hbar}{2M^{\ast}}\nabla^{2}+2%
{\displaystyle\sum\limits_{n}}
|\tilde{C}_{0n}|^{2}\omega_{0}(R_{nm})+2%
{\displaystyle\sum\limits_{n}}
|\tilde{C}_{0n}|^{2}\omega_{0,ac}(R_{nm})]\tilde{C}_{0m}(t)=0
\label{eq.:dC_o(x,y,t)/dt3(Ac+LO)2}%
\end{equation}
for variables $\tilde{C}_{0m}=C_{0m}\exp(i\bar{W}_{0}t).$ Using
Eq.(\ref{eq:local_limit}), we get%
\begin{equation}
i\frac{d}{dt}\tilde{C}_{0m}(t)+[\frac{\hbar}{2M^{\ast}}\nabla^{2}+2N|\tilde
{C}_{0m}|^{2}\omega_{0,ac}(k=0)+2 \sum_{n}
|\tilde{C}_{0n}|^{2}\omega_{0}(R_{nm})]\tilde{C}_{0m}(t)=0
\label{eq.:dC_om(t)/dt3(LO+ac)}%
\end{equation}
where the acoustic and LO phonons contribute to the local and nonlocal
nonlinear term, respectively.

\begin{equation}
i\frac{\partial }{\partial t}\tilde{C}_{0}(x,y,t)+[\frac{\hbar }{2M^{\ast }}%
\nabla ^{2}+2N|\tilde{C}_{0}(x,y,t)|^{2}\omega _{0,ac}(k=0)+\frac{2l}{%
l_{0}^{3}}\int \int |\tilde{C}_{0}(x^{\prime },y^{\prime },t)|^{2}\omega
_{0}(\sqrt{(x^{\prime }-x)^{2}+(y^{\prime }-y)^{2}})dx^{\prime }dy^{\prime }]%
\tilde{C}_{0}(x,y,t)=0  \label{eq.:dC_o(x,y,t)/dt(LO+ac)}
\end{equation}%
In the limit of weakly nonlocal NLS equation (see Section \ref%
{Section:weakly nonlocal}), Eq.(\ref{eq.:dC_o(x,y,t)/dt3}) is generalized as
\begin{equation}
i\frac{\partial }{\partial t}\bar{C}_{0}(\bar{x},\bar{y},t)+\frac{1}{2}%
\nabla ^{2}\bar{C}_{0}(\bar{x},\bar{y},t)+2N|\bar{C}_{0}(\bar{x},\bar{y}%
,t)|^{2}\omega _{0,ac}(k=0)\bar{C}_{0}(\bar{x},\bar{y},t)-\bar{J}\left[
\nabla ^{2}|\bar{C}_{0}(\bar{x},\bar{y},t)|^{2}\right] \bar{C}_{0}(\bar{x},%
\bar{y},t)=0  \label{eq.:dC_o(x,y,t)/dt3(Ac+LO)}
\end{equation}%
Equation (\ref{eq.:dC_o(x,y,t)/dt3(Ac+LO)}) corresponds to the "focusing" case
because $2N\omega_{0,ac}(k=0)>0$.
\end{widetext}

\subsection{The instability analysis for Eq.(\ref{eq.:dC_o(x,y,t)/dt3(Ac+LO)})}

Here we address plane wave solutions of the model
(\ref{eq.:dC_o(x,y,t)/dt3(Ac+LO)}) in the form of Eq.(\ref{eq:C_0(R,t)}).
Substituting expression (\ref{eq:C_0(R,t)}) in
Eq.(\ref{eq.:dC_o(x,y,t)/dt3(Ac+LO)}), we get%
\begin{equation}
\beta=\frac{\hbar}{2M^{\ast}}k_{0}^{2}-2N\rho_{0}\omega_{0,ac}(k=0)
\label{eq.:dispersion3}%
\end{equation}
In other words, the parameters satisfy the same dispersion relation as for the
standard local NLS equation. This is due to the absence of the "local"
nonlinear term for the Frohlich interaction.

Next, we carry out the linear stability analysis of the plane wave solutions,
in the form given by Eq.(\ref{eq:C_0(R,t)}). To this end, we assume as above
that $\tilde{C}_{0}(x,y,t)=[\sqrt{\rho_{0}}+d_{1}(\vec{\xi},t)]\exp
(i\mathbf{k}_{0}\cdot\mathbf{R}-i\beta t)$. Inserting
Eq.(\ref{eq:C_0(R,t)pert}) in the nonlocal NLS equation
(\ref{eq.:dC_o(x,y,t)/dt}) and linearizing around the solution
(\ref{eq:C_0(R,t)}) yields the evolution equation for the perturbation:%
\begin{widetext}
\begin{equation}
i\frac{\partial\bar{d}_{1}(\bar{\xi},t)}{\partial t}+\frac{1}{2}\nabla
_{\bar{\xi}}^{2}\bar{d}_{1}(\bar{\xi},t)+4N\rho_{0}\operatorname{Re}[d_{1}%
(\bar{\xi},t)] \omega_{0,ac}(k=0)+\frac{4l}{l_{0}^{3}}\rho_{0}\int
\operatorname{Re}[\bar{d}_{1}(\bar{\xi}^{\prime},t)]\omega_{0}(\bar{\xi
}^{\prime}-\bar{\xi})d\bar{\xi}^{\prime}=0 \label{eq:d_1(ksi,t)}%
\end{equation}
\end{widetext}
where we have switched to new coordinates $\bar{\xi}$ by analogy with
Section\ref{Section:solution1}. Decomposing the perturbation into real and
imaginary parts, $\bar{d}_{1}(\mathbf{\bar{\xi}},t)=u(\mathbf{\bar{\xi}%
},t)+iv(\mathbf{\bar{\xi}},t)$, we obtain two coupled equations for
$u(\mathbf{\bar{\xi}},t)$ and $v(\mathbf{\bar{\xi}},t)$. By introducing the
Fourier transforms, we reduce the linearized system to a set of ordinary
differential equations for Fourier transforms $u(\bar{k},t)$ and $v(\bar
{k},t)$, cf. Eq.(\ref{eq:Z}), with the only difference that $\omega_{0}%
(-\bar{k})$ in matrix $\hat{K}$, see Eq.(\ref{eq:Z,K}), should be replaced by
$\omega_{0}(-\bar{k})+\omega_{0,ac}(k=0)$. Accordingly, the eigenvalues
$\lambda$ of the matrix $\hat{K}$ are given by%
\begin{equation}
\lambda^{2}=-\bar{k}^{2}2N\rho_{0}\left(  \frac{1}{8N\rho_{0}}\bar{k}%
^{2}-\omega_{0}(-\bar{k})-\omega_{0,ac}(k=0)\right)  , \label{eq:lambda^2_ac}%
\end{equation}
and we get $\lambda^{2}<0$ (stability) when
\begin{equation}
\rho_{0}<\frac{1}{8N[\omega_{0}(-\bar{k})+\omega_{0,ac}(k=0)]}\bar{k}^{2}
\label{eq:stability_ac}%
\end{equation}
where both $\omega_{0}(-\bar{k})$ and $\omega_{0,ac}(k=0)$ are positive, and
according to Eq.(\ref{eq:omega_0,ac(k)}),%
\begin{equation}
\omega_{0,ac}(k=0)=\frac{2(C_{h}-C_{c})^{2}}{9\hbar NM\bar{u}^{2}}.
\label{eq:omega_0,ac(k=0)2}%
\end{equation}
It is easy to see that the exciton interaction with the acoustic phonons
reduces the intensity of modulationally stable waves. When wave number
$\bar{k}=0$, $\rho_{0}$ is also zero. When the exciton interaction with the
acoustical phonons is weak enough, $\omega_{0}(-\bar{k})$ may exceed
$\omega_{0,ac}(k=0)$ even in the range of small $\bar{k}$ where $\omega
_{0}(-\bar{k})\sim\bar{k}^{2}$. In that case we arrive at the same the
critical value $\rho_{cr}$, see Eq.(\ref{eq:(rho_0)_cr}).

The peculirities of Eq.(\ref{eq.:dC_o(x,y,t)/dt3(Ac+LO)}) in the weak
nonlocality limit, compared to the equations from
Refs.\cite{Krolikowski2001,Garcia_Malomed2003}, involve a reversal in the sign
of the nonlocal nonlinearity and an altered relationship between local and
nonlocal nonlinearities. This distinction gives rise to the difference between
the local and nonlocal dynamical regimes.

A particular instance of the plane wave solution, where $k_{0}=0$ and
$\beta=-2N\rho_{0}\omega_{0,ac}(k=0)$, represents a homogeneous distribution
that encompasses the superradiant state. Therefore, Eq.(\ref{eq:stability_ac})
is particularly significant for assessing the stability of the superradiant
state in the quasi-2D structures under the consideration, especially in the
context of the interaction of the excitons with both the LO and acoustic phonons.

\section{Nonlocal NLS equation in the polar coordinates and its fundamental
solution}

\label{Section:fundamental soliton}

Using the radial symmetry and the approximation of function $\omega_{0}(R)$
given in Appendix B, nonlocal equation (\ref{eq.:dC_o(x,y,t)/dt}) can be
written in the polar coordinates as%

\begin{widetext}
\begin{equation}
i\frac{\partial }{\partial t}\tilde{C}_{0}(R,\varphi ,t)+\frac{\hbar }{%
2M^{\ast }}\left[ \frac{1}{R}\frac{\partial }{\partial R}\left( R\frac{%
\partial }{\partial R}\right) +\frac{1}{R^{2}}\frac{\partial ^{2}}{\partial
\varphi ^{2}}\right] \tilde{C}_{0}(R,\varphi ,t)+\frac{2lG_{0}}{l_{0}^{3}}%
\tilde{C}_{0}(R,\varphi ,t)\exp \left( -\frac{R^{2}}{\rho ^{2}}\right)
\int_{0}^{\infty }dR^{\prime }R^{\prime }\left\vert \tilde{C}_{0}(R^{\prime
},t)\right\vert ^{2}g(R,R^{\prime })=0  \label{eq.:dC_0(R,t)/dt}
\end{equation}%
Here we use the notation%
\begin{equation}
g(R,R^{\prime })=2\pi \exp \left( -\frac{R^{\prime 2}}{\rho ^{2}}\right) %
\left[ \left( 1-\frac{R^{\prime 2}+R^{2}}{\rho ^{2}}\right) I_{0}\left(
\frac{2R^{\prime }R}{\rho ^{2}}\right) +\frac{2R^{\prime }R}{\rho ^{2}}%
I_{1}\left( \frac{2R^{\prime }R}{\rho ^{2}}\right) \right]
\label{eq:g(R,R')}
\end{equation}%
where $I_{n}(A)$ denotes the modified Bessel functions of the first kind.

Next, we move on to dimensionless variables $\tilde{R}=R/D$, $\tilde{t}%
=t/\tau $, $\tilde{\rho}=\rho /D$, and introduce the following vortex ansatz:%
\begin{equation}
\tilde{C}_{0}(\tilde{R},\varphi ,\tilde{t})=C(\tilde{R})\exp (im\varphi
)\exp (i\tilde{\Omega}\tilde{t}).  \label{eq:C(R)}
\end{equation}%
Substituting it into the dimensionless version of Eq. (\ref{eq.:dC_0(R,t)/dt}%
), we obtain the following equation for the radial amplitude $C(\tilde{R})$
(soliton profile):%
\begin{equation}
-\tilde{\Omega}C(\tilde{R})+\left[ \frac{1}{\tilde{R}}\frac{\partial }{%
\partial \tilde{R}}\left( \tilde{R}\frac{\partial }{\partial \tilde{R}}%
\right) -\frac{m^{2}}{\tilde{R}^{2}}\right] C(\tilde{R})+C(\tilde{R})\exp
\left( -\frac{\tilde{R}^{2}}{\tilde{\rho}^{2}}\right) \int_{0}^{\infty }d%
\tilde{R}^{\prime }\tilde{R}^{\prime }\left\vert C(\tilde{R}^{\prime
})\right\vert ^{2}g(\tilde{R},\tilde{R}^{\prime })=0.  \label{eq:C(R)2}
\end{equation}
\end{widetext}

\begin{figure*}
\begin{center}
\includegraphics[scale=0.4]{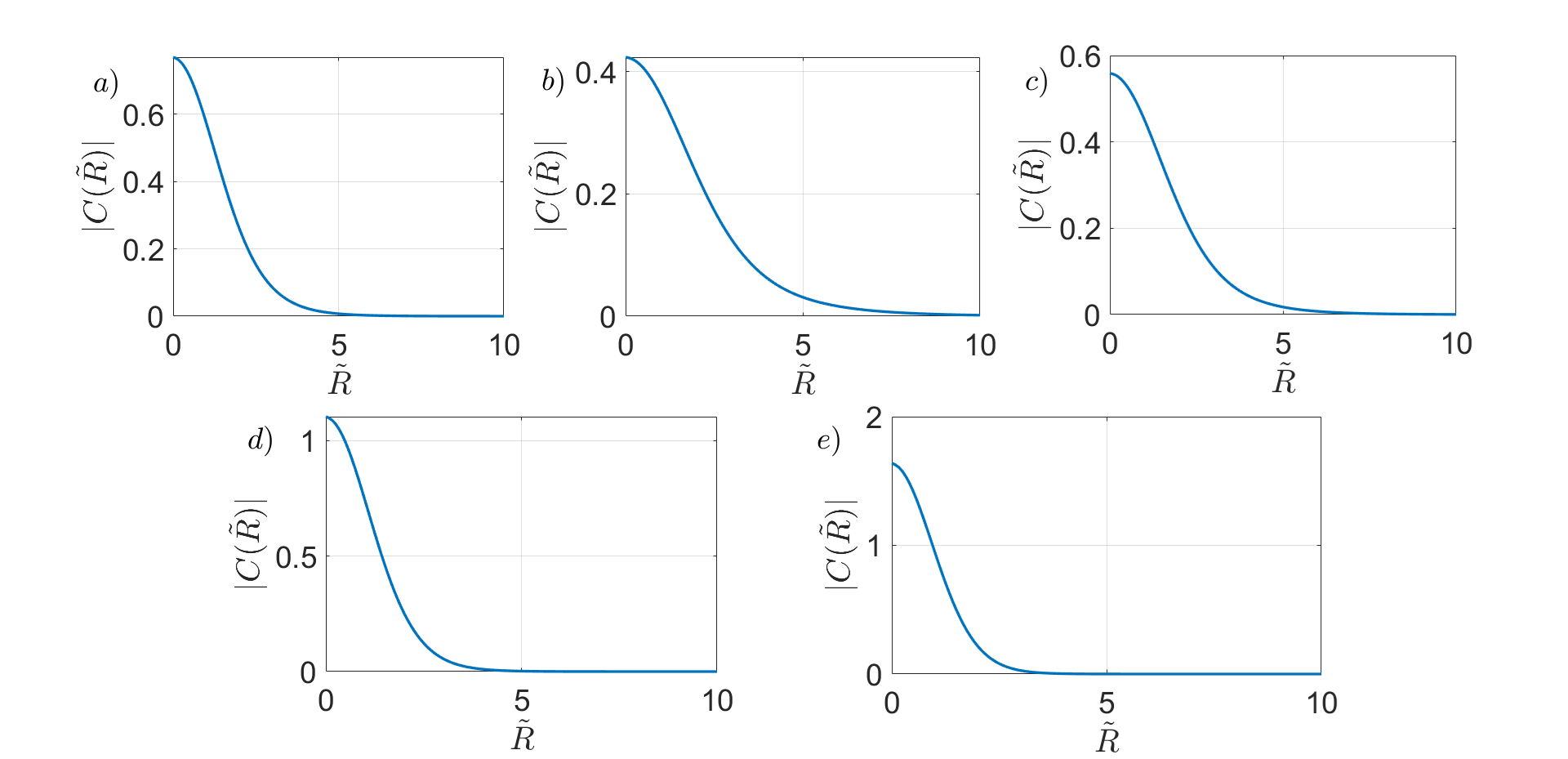}
\caption{\label{fig:profile}Fundamental soliton profile $C(\tilde{R})$ for $\tilde{\Omega}=1$
(a), $\tilde{\Omega}=0.25$ (b), $\tilde{\Omega}=0.5$ (c), $\tilde{\Omega}=2$
(d) and $\tilde{\Omega}=4$ (e). $R=\tilde{R}D$.}
\end{center}
\end{figure*}

The Newton's method was used to find the soliton profile. The radial
profile was computed over the interval from $\tilde{R}_{0}=1\times 10^{-5}$
(inner radius) to $\tilde{R}_{N}=20$ (outer radius), using $NUM_{R}=121$
(grid size) with radial step $d\tilde{R}=0.1(6)$ (radial grid step). The
Newton iteration was terminated when the convergence criterion $\max
\left\vert A_{n+1}-A_{n}\right\vert <10^{-14}$ was satisfied, where $A_{n}$
denotes the profile at the $n$-th iteration. Evaluation of the integral on
the left-hand side of Eq.(\ref{eq:C(R)2}) was performed using the
trapezoidal rule. Figure
\ref{fig:profile} shows profiles of the fundamental soliton for different
values of $\tilde{\Omega}$.%

Next, we carry out the linear stability analysis of the fundamental soliton
solution to the dimensionless version of Eq. (\ref{eq.:dC_0(R,t)/dt}). Assume
that%
\begin{equation}
\tilde{C}_{0}(\tilde{R},\varphi,\tilde{t})=[C(\tilde{R})+\delta\tilde
{C}(\tilde{R},\varphi,\tilde{t})]\exp(i\tilde{\Omega}\tilde{t})
\label{eq:C(R)pert}%
\end{equation}
where $\delta\tilde{C}(\tilde{R},\varphi,\tilde{t})=a(\tilde{R})\exp(i\delta
m\varphi+\tilde{\gamma}\tilde{t})+b^{\ast}(\tilde{R})\exp(-i\delta
m\varphi+\tilde{\gamma}^{\ast}\tilde{t})$ is the complex expression for the
small perturbation, and $C(\tilde{R})$ is the solution of Eq.(\ref{eq:C(R)2}).
The insertion of Eq.(\ref{eq:C(R)pert}) into the nonlocal NLS equation
(\ref{eq.:dC_0(R,t)/dt}) and linearization around solution (\ref{eq:C_0(R,t)})
yields the following evolution equations for the perturbation:%
\begin{widetext}
\begin{eqnarray}
\tilde{\gamma}a(\tilde{R})=-i\tilde{\Omega}a(\tilde{R})+\frac{i}{\tilde{R}}%
\frac{\partial }{\partial \tilde{R}}\left( \tilde{R}\frac{\partial a(\tilde{R%
})}{\partial \tilde{R}}\right) -i\frac{(m+\delta m)^{2}}{\tilde{R}^{2}}a(%
\tilde{R}) &+&iC(\tilde{R})\exp \left( -\frac{\tilde{R}^{2}}{\tilde{\rho}^{2}%
}\right)   \nonumber \\
\times \int_{0}^{\infty }d\tilde{R}^{\prime }\tilde{R}^{\prime }C(\tilde{R}%
^{\prime })[a(\tilde{R}^{\prime })+b(\tilde{R})]g(\tilde{R},\tilde{R}%
^{\prime })+ia(\tilde{R})\exp \left( -\frac{\tilde{R}^{2}}{\tilde{\rho}%
^{2}}\right) \int_{0}^{\infty }d\tilde{R}^{\prime }\tilde{R}^{\prime }|C(%
\tilde{R}^{\prime })|^{2}g(\tilde{R},\tilde{R}^{\prime })  \label{eq:a(R)a}
\end{eqnarray}%
\begin{eqnarray}
\tilde{\gamma}b(\tilde{R})=i\tilde{\Omega}b(\tilde{R})-\frac{i}{\tilde{R}}%
\frac{\partial }{\partial \tilde{R}}\left( \tilde{R}\frac{\partial b(\tilde{R%
})}{\partial \tilde{R}}\right) +i\frac{(m-\delta m)^{2}}{\tilde{R}^{2}}b(%
\tilde{R})-iC(\tilde{R})\exp \left( -\frac{\tilde{R}^{2}}{\tilde{\rho}^{2}%
}\right)   \nonumber \\
\times \int_{0}^{\infty }d\tilde{R}^{\prime }\tilde{R}^{\prime }C(\tilde{R}%
^{\prime })\left[ a(\tilde{R}^{\prime })+b(\tilde{R})\right] g(\tilde{R},%
\tilde{R}^{\prime })-ib(\tilde{R})\exp \left( -\frac{\tilde{R}^{2}}{\tilde{%
\rho}^{2}}\right) \int_{0}^{\infty }d\tilde{R}^{\prime }\tilde{R}^{\prime
}|C(\tilde{R}^{\prime })|^{2}g(\tilde{R},\tilde{R}^{\prime })
\label{eq:b(R)a}
\end{eqnarray}
\end{widetext}

\begin{figure}
\begin{center}
\includegraphics[
scale=0.33]{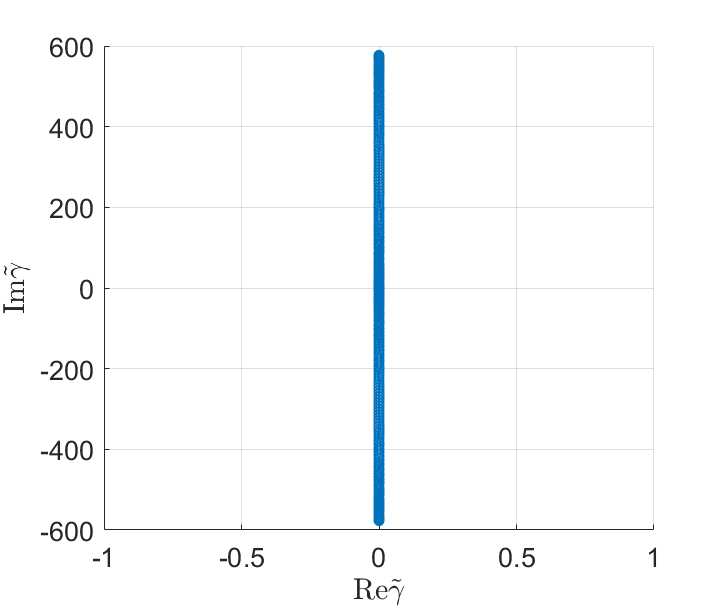}%
\caption{\label{fig:spectrum_gamma_dm=0}Spectrum of $\tilde{\gamma}$ for $\tilde{\Omega}=1$ and $\delta
m=0$.}
\end{center}
\end{figure}

\begin{figure*}
\begin{center}
\includegraphics[scale=0.33]{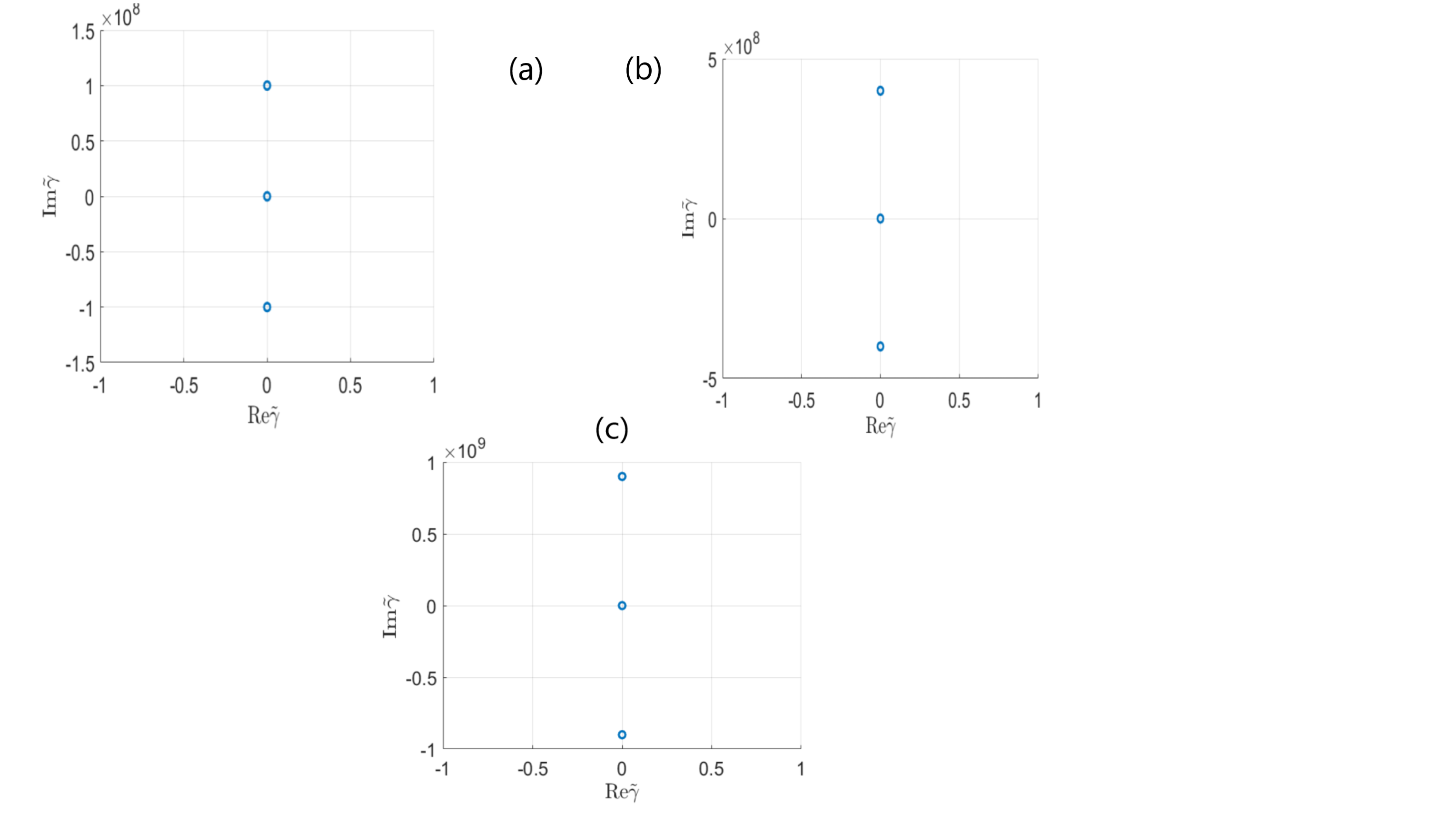}%
\caption{\label{fig:spectrum_gamma_dm>0}Spectra of $\tilde{\gamma}$ for $\tilde{\Omega}=1$ and $\delta m=1$
(a), $\delta m=2$ (b), $\delta m=3$ (c).}
\end{center}
\end{figure*}

We conducted simulations of the soliton evolution. Figures \ref{fig:spectrum_gamma_dm=0} and \ref{fig:spectrum_gamma_dm>0}
illustrate the stability of the fundamental soliton. Figure
\ref{fig:spectrum_gamma_dm=0} shows the spectrum of $\tilde{\gamma}$ for
$\tilde{\Omega}=1$ and $\delta m=0$.
Spectra of $\tilde{\gamma}$ for $\delta m=0$ and other $\tilde{\Omega}$ values
( $\tilde{\Omega}=0.25$, $0.5$, $2$ and $4$) {}{}are similar to the one
plotted in Fig. \ref{fig:spectrum_gamma_dm=0}.

Figure \ref{fig:spectrum_gamma_dm>0} shows the $\tilde{\gamma}$ spectra for
$\tilde{\Omega}=1$ and non-zero values of $\delta m$.
The $\tilde{\gamma}$ spectra for non-zero values of $\delta m$ and other
$\tilde{\Omega}$ values given in the Appendix D are similar. In all cases, the
real part of gamma is zero, which corresponds to a stable solution. The
spatial distribution of the fundamental soliton amplitude $|\tilde{C}%
_{0}(\tilde{x},\tilde{y},t)|$ at different times is shown in Appendix E.

Figures \ref{fig:spectrum_gamma_dm=0} and \ref{fig:spectrum_gamma_dm>0} of the
main text, and \ref{fig:spectrum_gamma_dm>0_Om=0.25},
\ref{fig:spectrum_gamma_dm>0_Om=4}, \ref{fig:amplitude} of Appendicies
illustrate the stability of the fundamental soliton, whose amplitude is
greater than the amplitude of the stable plane wave and the corresponding
superradiant state. In other words, the soliton regime increases the amplitude
of a stable exciton state that contributes to the development of the
superradiation though reduces the area in which the superradiant state is
present. One can evaluate the radius of this area. Indeed, according to
Fig.\ref{fig:profile}, the level of $C(\tilde{R})=0.1$, corresponding to the
critical value $\sqrt{\rho_{cr,0}}=0.1$ for plane waves, is achieved at
$R\approx3D=26.7\cdot10^{\mathbf{-}8}cm\approx a_{0}/2$, which corresponds to
half the exciton radius ($a_{0}\simeq55\mathring{A}$). Thus, in the case of
the formation of the fundamental soliton, the entire Wannier exciton with an
enhanced exciton amplitude can be involved in the process of SF. Increasing
the area of the superradiant state can be achieved in the case of the vortex
solitons, since they should lead to the formation of the corresponding
superradiant circular regions in thin films. This issue will be studied elsewhere.

\section{Conclusion}

\label{Section:conclusion}

In this work we have studied the stability of the superradiant state with
respect to the exciton-phonon interactions in hybrid perovskite thin films.
First, we considered the quasi-2D Wannier exciton interacting with the LO
phonons in polar crystals. We have derived the nonlinear equation governing
the coefficient of the exciton wavefunction in the single-exciton basis for
the lowest exciton state ($n=0$), or equivalently, the complex-valued
polarization, in the coordinate space. The obtained equation takes the form of
the 2D nonlocal NLS equation, which is well known in the theory of solitons.
We carried out the linear stability analysis of the plane wave solutions of
the equation under the consideration, and derived the stability criterion. We
have shown that the long-plane-wave solution is modulationally stable if its
squared amplitude $\rho_{0}$ does not exceed some critical value of the
intensity, which is expressed in terms of the exciton-LO
(longitudinal-optical) phonon interaction parameters.

Next, we explored the limit of the weakly nonlocal NLS equations. In this
scenario, the NLS equation transforms into a purely nonlocal form, which, to
the best of our knowledge, has not been previously addressed in the 2D
context. We have carried out the linear stability analysis of the plane wave
solutions of the latter equation and obtained results consistent with those
produced above for our general nonlinear equation in the long wave limit.

We also moved beyond the Fr\"{o}hlich picture and considered the interplay
between short-range (exciton-acoustic phonon) and long-range (exciton-LO
phonon) lattice coupling in the systems under the consideration. We extended
the nonlinear equations to the interaction with both the LO \ and acoustic
phonons. It has been shown that the nonlinear term for\ the exciton-acoustic
phonon interaction becomes local in contrast to the purely nonlocal term for
the interaction with the LO phonons. We have carried out the linear stability
analysis of the plane wave solutions for the case under the consideration and
shown that the exciton interaction with the acoustic phonons reduces the
intensity of the modulationally stable waves.

A particular instance of the plane wave solution represents the homogeneous
distribution that encompasses the superradiant state. Consequently, the
stability analysis carried out in this work is particularly significant for
assessing the stability of the superradiant state in the quasi-2D structures
under the consideration. The analytical results obtained in the work are
confirmed by numerical calculations.

We also numerically solved the 2D nonlocal NLS equation in the polar
coordinates and obtained its fundamental soliton solution. We studied its
stability and showed that the soliton regime increases the amplitude of the
stable exciton state that contributes to the development of SF though reduces
the area in which the superradiant state is present.

Extending the analysis, we plan to explore bright vortex solitons in the
present systems. The study of such solitons is certainly intriguing, since
they should lead to the formation of the corresponding superradiant circular
regions in thin films, increasing the area of the superradiant state in
comparison to the fundamental soliton. This consideration also brings dark
solitons into the focus. When dark solitons are generated, the localization of
the superradiant state is significantly reduced.

In conclusion, it should be noted that a recent experimental study \cite{Gundogdu2025Nature} in which the soliton concept was used to explain high-temperature SF in perovskites provides experimental supports for the theoretical considerations presented in our work.

\begin{acknowledgments}
B.F. acknowledges support by the European Cooperation in Science and
Technology (COST Action CA24109).
\end{acknowledgments}
%

\vspace{5mm}
\textbf{Author Contributions}. \textbf{A.A. Gladkij}: Formal analysis, Software, Validation, Visualization.
\textbf{N.A. Veretenov}: Software. \textbf{N.N. Rosanov:} Conceptualization,
Formal analysis, Methodology, Software, Supervision, Validation,
Visualization. \textbf{B.A. Malomed}: Conceptualization, Formal analysis,
Methodology, Writing -- review \& editing. \textbf{V.Al. Osipov}: Formal
analysis, Writing -- review \& editing. \textbf{B.D. Fainberg}:
Conceptualization, Formal analysis, Methodology, Project administration,
Validation, Visualization, Writing -- original draft, Writing -- review \&
editing.

\appendix

\section{Derivation of the equations for $\sigma_{\mathbf{k}}$
and $C_{n}(\mathbf{q}|t)$ using Heisenberg equations of motion}

The Hamiltonian for a Wannier exciton interacting with phonons typically
consists of three components: the Wannier exciton energy operator ($\hat
{H}_{ex}^{W}$), the phonon energy operator ($\hat{H}_{ph}$), and the Wannier
exciton-phonon interaction energy operator ($\hat{H}_{I}^{W}$)
\begin{equation}
\hat{H}_{0}=\hat{H}_{ex}^{W}+\hat{H}_{ph}+\hat{H}_{I}^{W} \label{eq.:H_03}%
\end{equation}
where
\begin{equation}
\hat{H}_{ex}^{W}=\hbar\sum_{n\mathbf{k}}W_{n}(\mathbf{k})B_{n\mathbf{k}}%
^{\dag}B_{n\mathbf{k}}, \label{eq:Wex_hamilt4}%
\end{equation}%
\begin{equation}
\hat{H}_{ph}=\sum_{\mathbf{\mathbf{q}}}\hbar\omega(\mathbf{q})b_{\mathbf{q}%
}^{\dag}b_{\mathbf{q}}, \label{eq.:H_ph}%
\end{equation}
\begin{widetext}
$b_{\mathbf{q}s}^{\dag}$($b_{\mathbf{q}s}$) are creation (annihilation)
operators of phonons,
\begin{equation}
\hat{H}_{I}^{W}=-i\hbar\sum_{nn^{\prime}\mathbf{kq}}D_{l}^{pol}B_{n,\mathbf{q}%
+\mathbf{k}}^{\dag}B_{n^{\prime}\mathbf{q}}(b_{\mathbf{k}}-b_{\mathbf{-k}%
}^{\dag}). \label{eq.:H^pol(W)_Id}%
\end{equation}

The Heisenberg equations of motion for operators $b_{\mathbf{q}}$ and
$B_{n\mathbf{q}}$ are as follows:%
\begin{equation}
\dot{b}_{\mathbf{q}}=\frac{i}{\hbar}\left[  \hat{H}_{0},b_{\mathbf{q}}\right]
=-i\omega(\mathbf{q})b_{\mathbf{q}}+\sum_{n^{\prime\prime}n^{\prime}}%
D_{l}^{pol}(\mathbf{-q};n^{\prime\prime}n^{\prime})\sum_{\mathbf{k}%
}B_{n^{\prime\prime},\mathbf{k}}^{\dag}B_{n^{\prime},\mathbf{k+q}}
\label{eq:db_q/dt}%
\end{equation}

\begin{equation}
\dot{B}_{n\mathbf{q}}=\frac{i}{\hbar}\left[  \hat{H}_{0},B_{n\mathbf{q}%
}\right]  =-iW_{n}(\mathbf{q})B_{n\mathbf{q}}+\sum_{n^{\prime}\mathbf{k}}%
D_{l}^{pol\ast}(\mathbf{k};n^{\prime}n)B_{n^{\prime}\mathbf{q+k}%
}(b_{\mathbf{k}}^{\dag}-b_{\mathbf{-k}}) \label{eq:dB_nk/dt}%
\end{equation}
\end{widetext}
Averaging these equations using the wave function, Eq.(\ref{wavefunction1}),
together with the normalization condition, Eq.(\ref{eq:Normalization}), we
obtain Eq.(\ref{eq:dsigma_ks/dt}) from the main text for $\sigma_{\mathbf{k}}$
in the case $n^{\prime\prime}=n^{\prime}=0$ in Eq.(\ref{eq:db_q/dt}). In
Eq.(\ref{eq:dsigma_ks/dt}), we also introduce a small decay $\gamma$ rate of
the mode $\omega_{l}$.

From Eq.(\ref{eq:dB_nk/dt}) we obtain
\begin{equation}
\dot{C}_{n}(\mathbf{q}|t)=-iW_{n}(\mathbf{q})C_{n}(\mathbf{q}|t)+\sum
_{n^{\prime}\mathbf{k}}\tilde{\alpha}_{l}\mathbf{(k};n^{\prime}n)C_{n^{\prime
}}(\mathbf{q+k}|t) \label{eq:dC_n/dt}%
\end{equation}
when $\sum_{\mathbf{q}^{\prime}}G^{\ast}(\mathbf{q}^{\prime}|t)\approx1$. Here
$\tilde{\alpha}_{l}\mathbf{(k};n^{\prime}n)\mathbf{=}D_{l}^{pol\ast
}\mathbf{(k};n^{\prime}n)(\sigma_{\mathbf{k}}^{\ast}-\sigma_{\mathbf{-k}})$.
Equation (\ref{eq:dC_n/dt}) differs from Eq.(32) of
Ref.\cite{Fainberg_Osipov2024JCP} by omitting both an insignificant correction
to the exciton eigenenergy $\hbar W_{n}(\mathbf{q})$ and the term
$\mathbf{-}\hat{\Gamma}(C)$ on the right-hand side, which describes the
effective spontaneous decay of the exciton state resulting from its
interaction with the electromagnetic vacuum. The decay term $\hat{\Gamma}(C)$
at $n=0$ takes the simple form $\hat{\Gamma}(C)|_{n=0}=\frac{1}{2}%
\Gamma_{\mathbf{q}}(\omega)C_{0}(\mathbf{q}|t)$ where $\Gamma_{\mathbf{q}%
}(\omega)$ is given by Eq.(34) of Ref.\cite{Fainberg_Osipov2024JCP}. For the
superradiant state ($\mathbf{q}=0$) \cite{Fainberg_Osipov2024JCP}, the decay
rate is given by $\Gamma_{\mathbf{q}}(\omega)=\Gamma_{\mathbf{q}=0}%
(\omega)\equiv\Gamma_{0}(\omega)=24\pi\left(  \frac{\lambda_1}{a_{0}}\right)
^{2}\left(  \frac{\omega_{0}}{\omega}\right)  ^{2}\gamma_{s}$, where
$\gamma_{s}=\frac{4|\mathbf{D}_{cv}|^{2}}{3\hbar\lambda_1^{3}}$, $\lambda_1$ is
the wavelength corresponding to the optical transition from the lowest-energy
exciton state ($n=0$), and $\mathbf{D}_{cv}$ is the interband transition
dipole moment. The radiative decay rate is thus enhanced by a factor of
$24\pi\left(  \frac{\lambda_1}{a_{0}}\right)  ^{2}$. This enhancement arises
from the coherent nature of the two-dimensional exciton with respect to its
center-of-mass motion. If the dimensions of the coherent region are smaller
than the wavelength - which occurs, e.g., for a localized soliton - then, for
the sake of the estimate, the factor $\sim\left(  \lambda_1/a_{0}\right)  ^{2}$
should be replaced by $\sim\left(  \Delta x/a_{0}\right)  ^{2}$ where $\Delta
x$ is the characteristic size of the soliton.

\section{Approximation of function $\omega_{0}(R)$}

\begin{figure}
\begin{center}
\includegraphics[
scale=0.5]{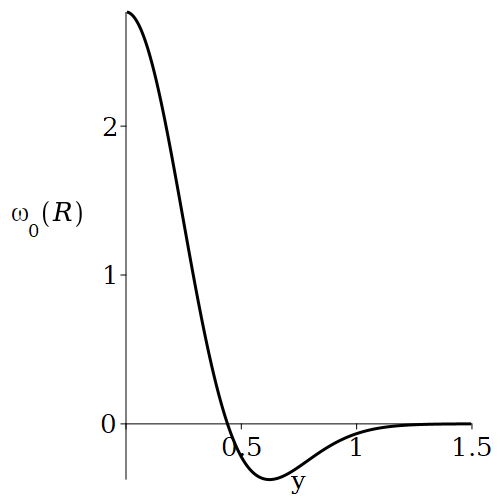}
\caption{\label{fig:gauss}Approximation of $\omega_{0}(R)$ using Eq.(\ref{eq:omega_0(R)appr}).
$y=R/a_{0}$.}
\end{center}
\end{figure}

For the function $\omega_{0}(R)$ we {}{}used the following approximation%

\begin{equation}
\omega_{0}(R)=G_{0}\exp(-R^{2}/\rho^{2})\left(  1-\frac{R^{2}}{r_{0}^{2}%
}\right)  , \label{eq:omega_0(R)appr}%
\end{equation}
where, according to Eq.(\ref{eq:omega_0(R_nm)}),
\begin{equation}
G_{0}=\omega_{0}(R=0)=\frac{L^{2}}{2\pi}\int_{0}^{\infty}dkk\omega_{0}(k)
\label{eq:G_0}%
\end{equation}
and $\rho$ and $r_{0}$ are fitting parameters. The integral in the last term
on the left-hand side of Eq.(\ref{eq:dispersion1}) can be written as $2\pi
\int_{0}^{\infty}dRR\omega_{0}(R)$ in the polar coordinates, which vanishes in
the case of Wannier excitons interacting with LO phonons. Therefore, $\int
_{0}^{\infty}dRR\omega_{0}(R)=0$ that leads to $r_{0}=\rho$. Since exact
function $\omega_{0}(R)=0$ at $R=0.45a_{0}$ (see Fig.\ref{omega_0(R_nm)}), we
get $\rho=0.45a_{0}$. Equation (\ref{eq:omega_0(R)appr}) for $r_{0}%
=\rho=0.45a_{0}$ (see Fig.\ref{fig:gauss}) reproduces the main features of the
parameter of the exciton-phonon interaction $\omega_{0}(R_{m^{\prime}m})$
shown in Fig.\ref{omega_0(R_nm)}.

\begin{widetext}
\section{Non-dimensionalization of Eq.(\ref{eq.:dC_o(x,y,t)/dt})}

Let $\tilde{x}=x/D$, $\tilde{y}=y/D$ and $\tilde{t}=t/\tau$ where $\tilde{x},$
$\tilde{y}$ and $\tilde{t}$ are dimensionless variables. Then we get from
Eq.(\ref{eq.:dC_o(x,y,t)/dt}) using the approximation based on
Eq.(\ref{eq:omega_0(R)appr}):%
\begin{eqnarray}
&  i\frac{\partial}{\partial\tilde{t}}\tilde{C}_{0}(\tilde{x},\tilde{y}%
,\tilde{t})+\frac{\hbar\tau}{2M^{\ast}D^{2}}(\frac{\partial^{2}}%
{\partial\tilde{x}^{2}}+\frac{\partial^{2}}{\partial\tilde{y}^{2}})\tilde
{C}_{0}(\tilde{x},\tilde{y},\tilde{t})+\frac{2l\tau G_{0}D^{2}}{l_{0}^{3}%
}\tilde{C}_{0}(\tilde{x},\tilde{y},\tilde{t})\int\int d\tilde{x}^{\prime
}d\tilde{y}^{\prime}|\tilde{C}_{0}(\tilde{x}^{\prime},\tilde{y}^{\prime
},\tilde{t})|^{2}\nonumber\\
&  \times\exp\left[  -\frac{D^{2}}{\rho^{2}}(\tilde{x}^{\prime}-\tilde{x}%
)^{2}\right]  \exp\left[  -\frac{D^{2}}{\rho^{2}}(\tilde{y}^{\prime}-\tilde
{y})^{2}\right]  \left[  1-\frac{D^{2}}{\rho^{2}}(\tilde{x}^{\prime}-\tilde
{x})^{2}-\frac{D^{2}}{\rho^{2}}(\tilde{y}^{\prime}-\tilde{y})^{2}\right]
  =0 \label{eq.:dC_o(x,y,t)/dt_dmnls}
\end{eqnarray}
\end{widetext}

Denoting $A=\frac{\hbar\tau}{2D^{2}M^{\ast}}$ and $B=\frac{2l\tau G_{0}D^{2}%
}{l_{0}^{3}}$, one can easily see that the product $AB$ does not depend on
$D$. We put $AB=1$. Then we obtain for the characteristic time
\begin{equation}
\tau=\sqrt{\frac{M^{\ast}l_{0}^{3}}{\hbar lG_{0}}} \label{eq:tau}%
\end{equation}
If in addition, we put $B=1$, we get for the characteristic length%
\begin{equation}
D=\sqrt[4]{\frac{\hbar l_{0}^{3}}{4M^{\ast}lG_{0}}} \label{eq:D}%
\end{equation}
The respective values in physical units {}{}are obtained by multiplying by the
characteristic length and time.

Bearing in mind that $\rho=0.45a_{0}$, one can evaluate parameter $\frac
{D}{\rho}\simeq0.32$ in dimensionless equation (\ref{eq.:dC_o(x,y,t)/dt_dmnls}).

\section{Spectra of $\tilde{\gamma}$ for non-zero values of
$\delta m$ and different $\tilde{\Omega}$ values.}

\begin{figure*}
\begin{center}
\includegraphics[scale=0.33]{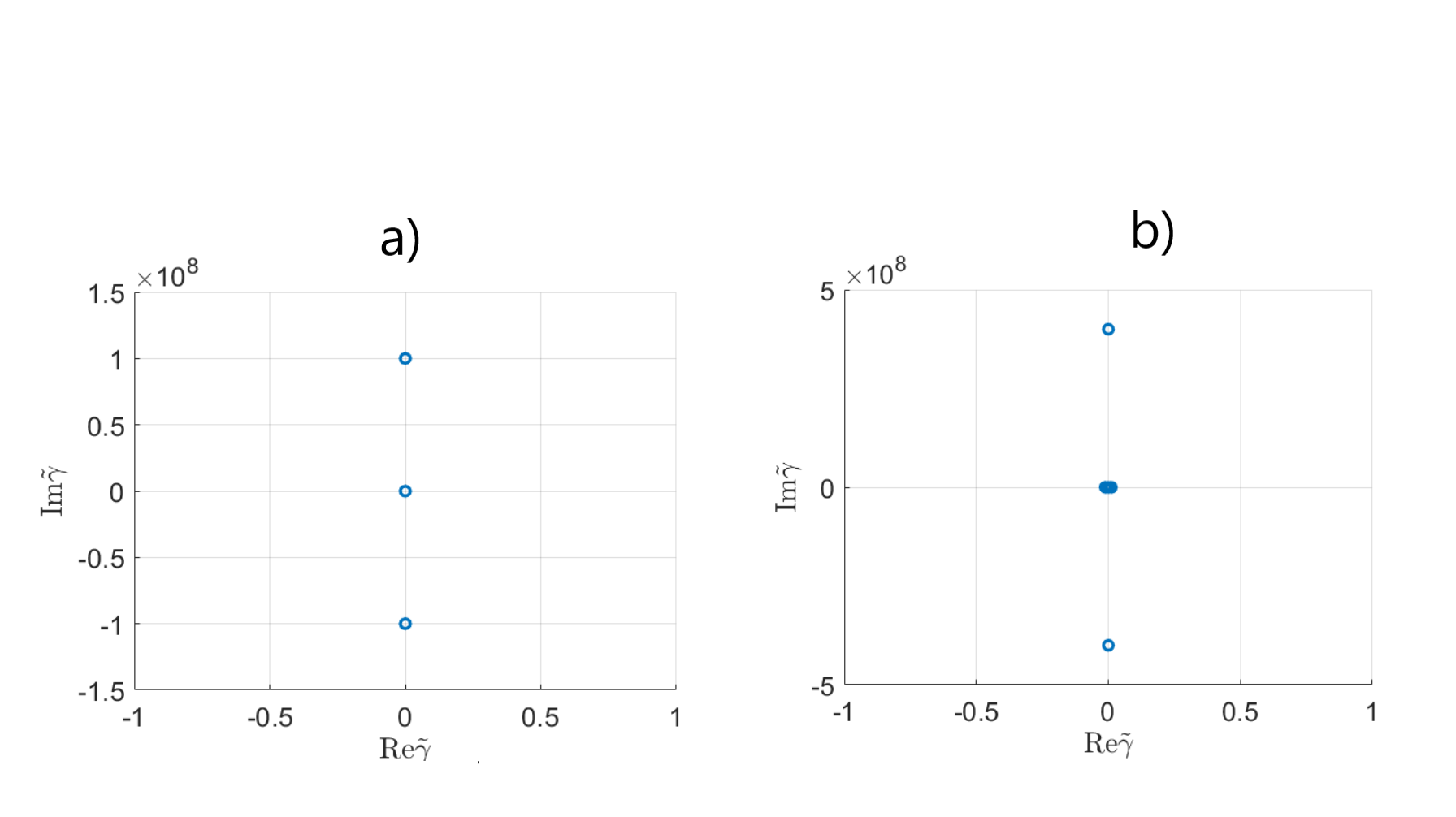}
\caption{Spectra of $\tilde{\gamma}$ for $\tilde{\Omega}=0.25$ and $\delta
m=1$ (a), $\delta m=2$ (b).}%
\label{fig:spectrum_gamma_dm>0_Om=0.25}%
\end{center}
\end{figure*}

\begin{figure*}
\begin{center}
\includegraphics[scale=0.33]{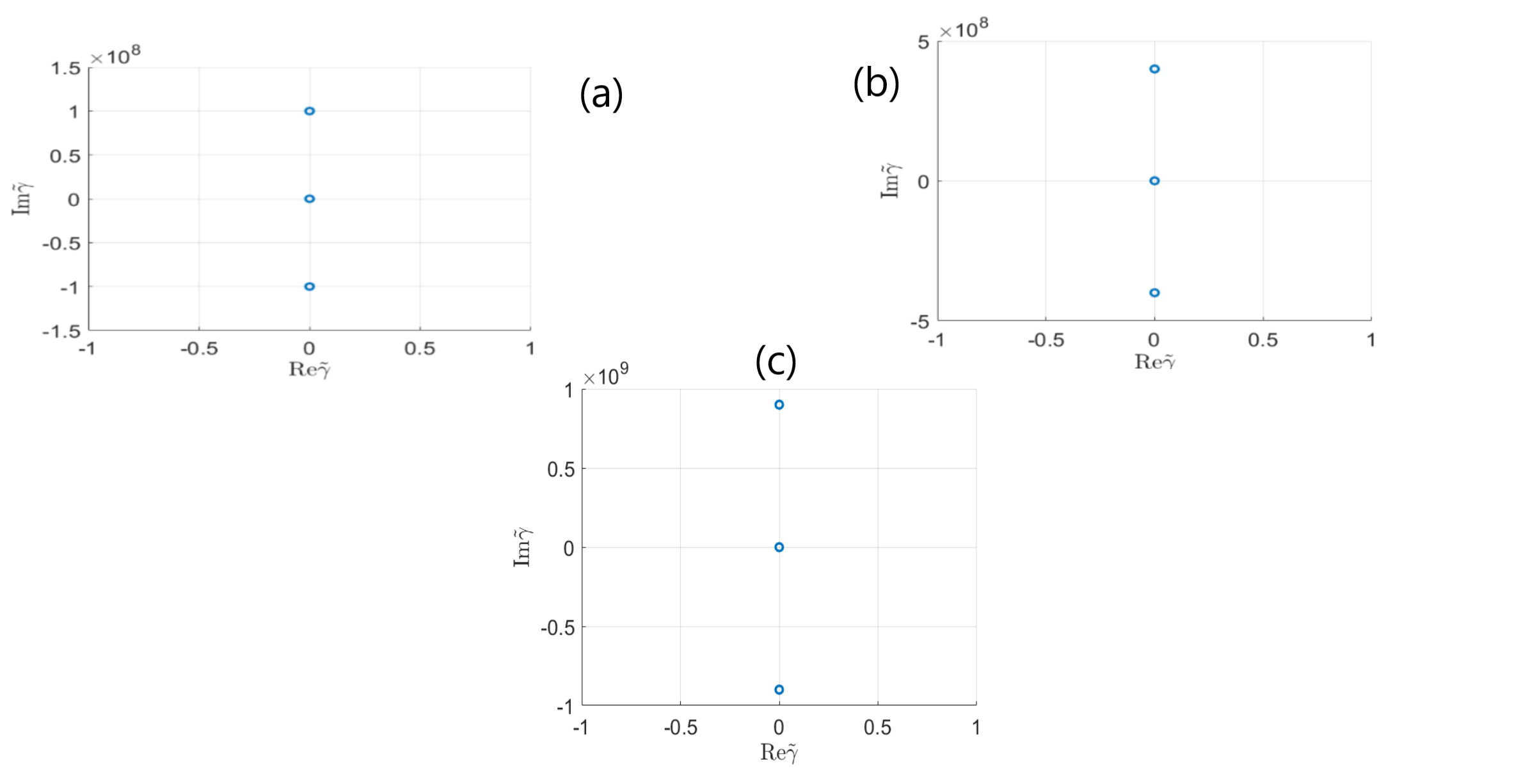}%
\caption{Spectra of $\tilde{\gamma}$ for $\tilde{\Omega}=4$ and $\delta m=1$
(a), $\delta m=2$ (b), $\delta m=3$ (c).}%
\label{fig:spectrum_gamma_dm>0_Om=4}%
\end{center}
\end{figure*}

Figures \ref{fig:spectrum_gamma_dm>0_Om=0.25} and
\ref{fig:spectrum_gamma_dm>0_Om=4} show the $\tilde{\gamma}$ spectra for
non-zero values of $\delta m$ and different $\tilde{\Omega}$. The real part of
gamma is zero, which corresponds to a stable solution.%


\section{Spatial shapes of the fundamental solitons at different
times}

Below one can see the spatial distribution of the fundamental soliton
amplitude $|\tilde{C}_{0}(\tilde{x},\tilde{y},t)|$ at different times.%
The computational domain was defined by $\tilde{x}_{\min }=y_{\min }=-14.1421,$ $%
\tilde{x}_{\max }=\tilde{y}_{\max }=13.4421$ with $NUM_{X}=NUM_{Y}=51$ (grid
size). The spatial step was $d\tilde{x}=d\tilde{y}=0.551685424949238$, and
the time step was $d\tilde{t}=0.01$. The simulations were carried out up to $%
NUM_{T}=3000$ time steps. Zero Dirichlet boundary conditions were imposed.
Time integration was performed using the fourth-order Runge-Kutta method.
The initial condition was taken as a soliton with a noise-perturbed
amplitude, $A(\tilde{x},\tilde{y})=A_{sol}(\tilde{x},\tilde{y})+0.01\cdot
noise$, where $A_{sol}(\tilde{x},\tilde{y})$ was obtained using the Newton
method and $noise=rand(\tilde{x},\tilde{y})$.
\begin{figure*}
\begin{center}
\includegraphics[scale=0.33]{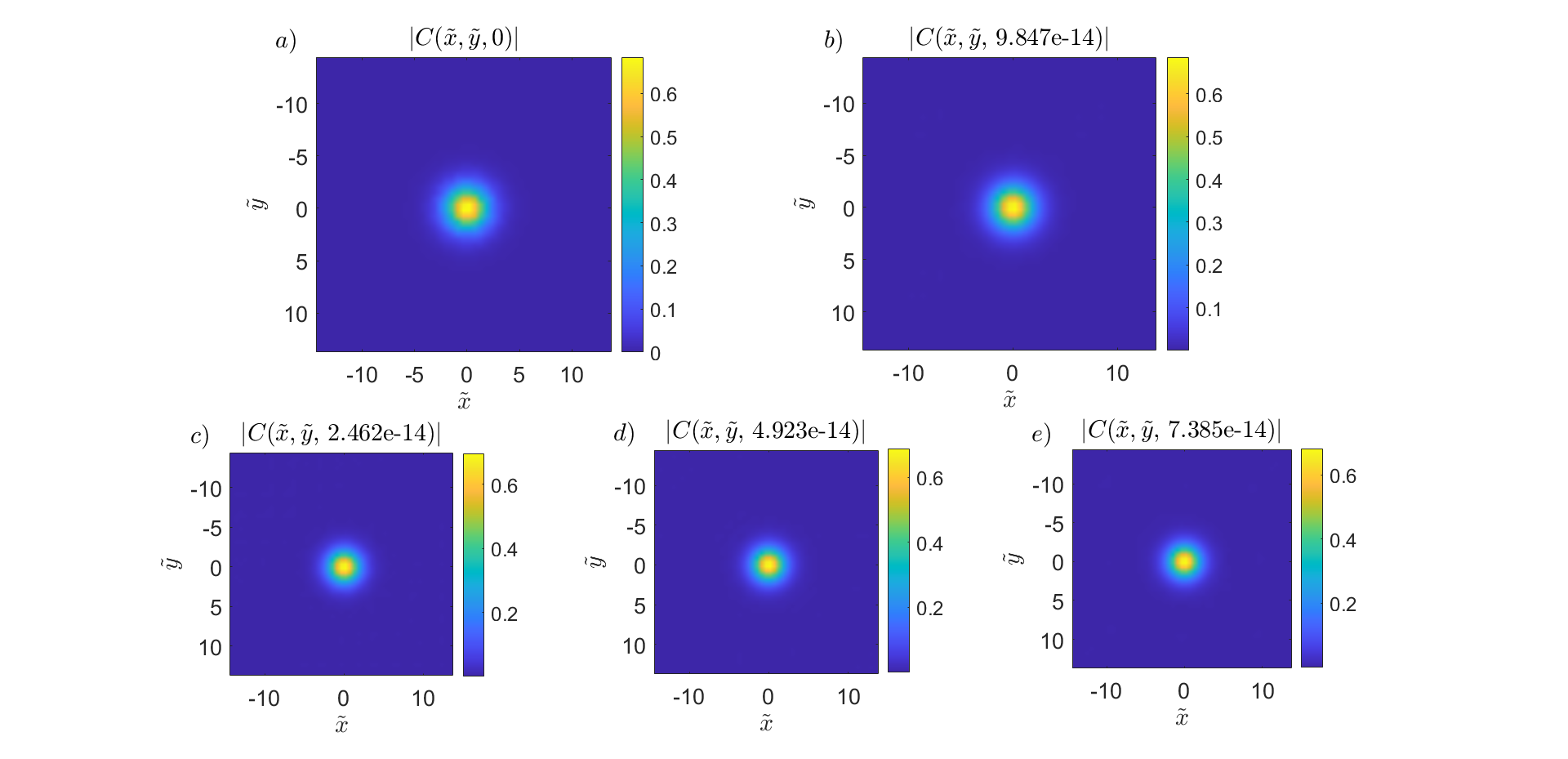}
\caption{Spatial distribution of the fundamental soliton amplitude $|\tilde
{C}_{0}(\tilde{x},\tilde{y},t)|$ for $\tilde{\Omega}=1$ at the initial moment
of time $t=0$ (a), $t=9.85\cdot10^{-14}$ $s$ (b), $t=2.46\cdot10^{-14}$ $s$
(c), $t=4.92\cdot10^{-14}$ $s$ (d) and $t=7.39\cdot10^{-14}$ $s$ (e).}%
\label{fig:amplitude}%
\end{center}
\end{figure*}

Figure \ref{fig:amplitude} illustrates the stability of the fundamental soliton.

\section{Newly introduced variables including dimensionless quantities}

All newly introduced variables, including the dimensionless quantities, are
summarized below. We introduced the scaled coordinates $\bar{x}=\sqrt{\frac{%
M^{\ast }}{\hbar }}x$, $\bar{y}=\sqrt{\frac{M^{\ast }}{\hbar }}y$, $\bar{R}=%
\sqrt{\frac{M^{\ast }}{\hbar }}\mathbf{R}$ and $\bar{\xi}=\sqrt{\frac{%
M^{\ast }}{\hbar }}\vec{\xi}$ , and the scaled wave number $\bar{k}=\sqrt{%
\frac{\hbar }{M^{\ast }}}k$. The dimensionless spatial coordinates and time
are defined as $\tilde{x}=x/D$, $\tilde{y}=y/D$, $\tilde{R}=R/D$ and $\tilde{%
t}=t/\tau $ where $\tau $ and $D$ are given by Eqs.(\ref{eq:tau}) and (\ref%
{eq:D}) of Appendix C, respectively. The dimensionless plane wave intensity $%
\rho _{0}$ is defined by Eq.(\ref{eq:rho_0}) of the main text as $\rho _{0}=%
\frac{1}{N}\left[ 1-\sum_{\mathbf{q}}|G(\mathbf{q}|t)|^{2}\right] $, where $%
N $ is the number of particles (electron-hole pairs) and $G(\mathbf{q}|t)$
is the ground state amplitude.




\clearpage



\begin{thebibliography}{10}
\providecommand{\url}[1]{#1}
\def\UrlFont{\rmfamily}
\providecommand{\newblock}{\relax}
\providecommand{\bibinfo}[2]{#2}
\providecommand\BIBentrySTDinterwordspacing{\spaceskip=0pt\relax}
\providecommand\BIBentryALTinterwordstretchfactor{4}
\providecommand\BIBentryALTinterwordspacing{\spaceskip=\fontdimen2\font plus
\BIBentryALTinterwordstretchfactor\fontdimen3\font minus
  \fontdimen4\font\relax}
\providecommand\BIBforeignlanguage[2]{{%
\expandafter\ifx\csname l@#1\endcsname\relax
\typeout{** WARNING: IEEEtran.bst: No hyphenation pattern has been}%
\typeout{** loaded for the language `#1'. Using the pattern for}%
\typeout{** the default language instead.}%
\else
\language=\csname l@#1\endcsname
\fi
#2}}

\bibitem{Gundogdu2021Nature_Phot}
G.~Findik, M.~Biliroglu, D.~Seyitliyev, J.~Mendes, A.~Barrette, H.~Ardekani,
  L.~Lei, Q.~Dong, F.~So, and K.~Gundogdu, High-temperature superfluorescence
  in methyl ammonium lead iodide, Nature Photonics \textbf{15},
  676 (2021).

\bibitem{Gundogdu2022Nature_Phot}
M.~Biliroglu, G.~Findik, J.~Mendes, D.~Seyitliyev, L.~Lei, Q.~Dong, Y.~Mehta,
  V.~V. Temnov, F.~So, and K.~Gundogdu, Room-temperature superfluorescence in
  hybrid perovskites and its origins, Nature Photonics \textbf{16}, 324 (2022).

\bibitem{Wildenborg2025ACSPhotonics}
A.~J. Wildenborg, R.~J. Munter, F.~Freire-Fernandez, E.~T.~F. Freitas, and
  J.~Y. Suh, Superlattice-induced superfluorescence in quasi-2d metal halide
  perovskites, ACS Photonics \textbf{12}, 3476 (2025).

\bibitem{Fainberg_Osipov2024JCP}
B.~D. Fainberg and V.~A. Osipov, Theory of high-temperature superfluorescence
  in hybrid perovskite thin films, J. Chem. Phys. \textbf{161}, 114705 (2024).

\bibitem{Long19Nanoscale}
H. Long, X. Peng, J. Lu, K. Lin, L. Xie, B. Zhang, L. Yinga, and Z. Wei, Exciton-phonon
  interaction in quasi-two dimensional layered (PEA)2(CsPbBr3)(n-1)PbBr4
  perovskite, Nanoscale \textbf{11}, 21867 (2019).

\bibitem{Malomed2022}
B.~A. Malomed, Two-dimensional solitons in nonlocal media: A brief review,
  Symmetry \textbf{14}, 1565 (2022).

\bibitem{Krolikowski2001}
W.~Krolikowski, O.~Bang, J.~J. Rasmussen, and J.~Wyller, Modulational
  instability in nonlocal nonlinear kerr media, Phys. Rev. E \textbf{64}, 016612 (2001).

\bibitem{Krolikowsk2004}
W.~Krolikowski, Modulational instability, solitons and beam propagation in
  spatially nonlocal nonlinear media, J. Opt. B: Quantum Semiclass.
  Opt. \textbf{6}, S288 (2004).

\bibitem{Silva2020JPCL}
A.~R.~S. Kandada and C.~Silva, Exciton polarons in two-dimensional hybrid
  metal-halide perovskites, J. Phys. Chem. Lett. \textbf{11}, 3173 (2020).

\bibitem{Glauber63}
R.~J. Glauber, Coherent and incoherent states of the radiation field,
  Phys. Rev \textbf{131}, 2766 (1963).

\bibitem{BBGK1971}
\BIBentryALTinterwordspacing
J.-M. Sixdeniers and K.~A. Penson, On the completeness of coherent states
  generated by binomial distribution, Journal of Physics A:
  Mathematical and General \textbf{33}, 2907 (2000).
  [Online]. Available: \url{https://doi.org/10.1088/0305-4470/33/14/319}
\BIBentrySTDinterwordspacing

\bibitem{Davydov82}
A.~S. Davydov, Solitons in quasi-one-dimensional molecular structures,
  Soviet Physics Uspekhi \textbf{25}, 898 (1982).

\bibitem{Gelin2015}
K.-W. Sun, M.~F. Gelin, V.~Y. Chernyak, and Y.~Zhao, Davydov ansatz as an
  efficient tool for the simulation of nonlinear optical response of molecular
  aggregates, J. Chem. Phys. \textbf{142}, 212448 (2015).

\bibitem{Kit63}
C.~Kittel, \emph{Quantum theory of solids}.\hskip 1em plus 0.5em minus
  0.4em\relax (New York-London, John Wiley and Sons, 1963).

\bibitem{Dav71}
A.~S. Davydov, \emph{Theory of Molecular Excitons}.\hskip 1em plus 0.5em minus
  0.4em\relax (New York, Plenum, 1971).

\bibitem{Agranovich09}
V.~M. Agranovich, \emph{Excitations in Organic Solids}.\hskip 1em plus 0.5em
  minus 0.4em\relax (New York, Oxford University Press, 2009).

\bibitem{Fainberg19JPCC}
B.~D. Fainberg, Study of electron-vibrational interaction in molecular
  aggregates using mean-field theory: From exciton absorption and luminescence
  to exciton-polariton dispersion in nanofibers, J. Phys. Chem. C
  \textbf{123}, 7366 (2019).

\bibitem{Krolikowski2000}
W.~Krolikowski and O.~Bang, Solitons in nonlocal nonlinear media: Exact
  solutions, Phys. Rev. E \textbf{63}, 016610 (2000).

\bibitem{Gundogdu2025Nature}
M.~Biliroglu, M.~Turel, A.~Ghita, M.~Kotyrov, X.~Qin, D.~Seyitliyev,
  N.~Phonthiptokun, M.~Abdelsamei, J.~Chai, R.~Su, U.~Herath, A.~K. Swan, V.~V.
  Temnov, V.~Blum, F.~So, and K.~Gundogdu, Unconventional solitonic
  high-temperature superfluorescence from perovskites, Nature \textbf{642}, 71 (2025).

\bibitem{Frohlich54}
H.~Frohlich, Electrons in lattice fields, Advances in Physics \textbf{3}, 325 (1954).

\bibitem{Feynman98}
R.~P. Feynman, \emph{Statistical Mechanics}.\hskip 1em plus 0.5em minus
  0.4em\relax (Boulder, Colorado, Westview press, 1998).

\bibitem{Jai_Singh94}
J.~Singh, \emph{Excitation Energy Transfer in Condensed Matter. Theory and
  Applications}.\hskip 1em plus 0.5em minus 0.4em\relax (New York, Springer
  Science+Business Media, 1994).

\bibitem{Miyata17}
K.~Miyata, D.~Meggiolaro, M.~T. Trinh, P.~P. Joshi, E.~Mosconi, S.~C. Jones,
  F.~D. Angelis, and X.-Y. Zhu, Large polarons in lead halide perovskites,
  Sci. Adv. \textbf{3}, e1701217 (2017).

\bibitem{Malomed2025}
W.-R. Sun and B.~A. Malomed, Subharmonic modulational instabilities,
  Phys. Rep. \textbf{1139}, 1 (2025).

\bibitem{Frohna18}
K.~Frohna, T.~Deshpande, J.~Harter, W.~Peng, B.~A. Barker, J.~B. Neaton, S.~G.
  Louie, O.~M. Bakr, D.~Hsieh, and M.~Bernardi, Inversion symmetry and bulk
  rashba effect in methylammonium lead iodide perovskite single crystals,
  Nature Communications \textbf{9}, 1829 (2018).

\bibitem{Porkolab1976}
M.~Porkolab and M.~V. Goldman, Upper-hybrid solitons and
  oscillating-two-stream instabilities, Phys. Fluids \textbf{19}, 872 (1976).

\bibitem{Garcia_Malomed2003}
J.~J. Garcia-Ripoll, V.~V. Konotop, B.~Malomed, and V.~M. Perez-Garcia, A
  quasi-local Gross-Pitaevskii equation for attractive bose-einstein
  condensates, Mathematics and Computers in Simulation \textbf{62},
  21 (2003).

\bibitem{Toyozawa59}
Y.~Toyozawa, On the dynamical behavior of an exciton, Progr. Theor.
  Phys. Suppl. \textbf{12}, 111 (1959).

\bibitem{Toyozawa58}
Y.~Toyozawa, Theory of line-shapes of the excit.on absorption bands,
  Progr. Theor. Phys. \textbf{20}, 53 (1958).

\end{thebibliography}
\end{document}